\documentclass[12pt,preprint]{aastex}

\begin{document}

\begin{center}
\large\textbf{Electromagnetic streaming instabilities of magnetized
accretion disks with strong collisional coupling of species}
\large \\

\smallskip
\smallskip \textsf{A. K. Nekrasov}
\end{center}

Institute of Physics of the Earth, Russian Academy of Sciences, 123995
Moscow, Russia

\noindent e-mail: anatoli.nekrassov@t-online.de

\smallskip

\noindent \textbf{Abstract. }Electromagnetic streaming instabilities of
multicomponent collisional magnetized accretion disks are studied.
Sufficiently ionized regions of the disk are explored where there is strong
collisional coupling of neutral atoms with both ions and dust grains
simultaneously. The steady state is investigated in detail and the azimuthal
and radial background velocities of species are calculated. The azimuthal
velocity of ions, dust grains, and neutrals is found to be less than the
Keplerian velocity. The radial velocity of neutrals and dust grains is shown
to be directed inward of the disk. The general solution for the perturbed
velocities of species taking into account collisions and thermal pressure is
obtained. The effect on the collisional frequencies, due to density
perturbations of charged species and neutrals, is included. It is shown that
dust grains can be involved in the fast electromagnetic perturbations
induced by the ions and electrons through the strong collisions of these
grains with neutrals that in turn have a strong collisional coupling with
the ions. The dispersion relation for the vertical perturbations is derived
and its unstable solutions due to different background velocities of ions
and electrons are found. The growth rates of the streaming instabilities
considered can be much larger than the Keplerian frequency.

\noindent \textit{Subject headings: }accretion, accretion
disks instabilities, magnetic fields, plasma-waves

\section{Introduction}

Protostellar and protoplanetary accretion disks are known to contain
submicron- and micron-sized solid particles (dust grains) (e.g., Beckwith \&
Sargent 1996; Isella et al. 2006; Besla \& Wu 2007; Quanz et al. 2007; Pinte
et al. 2007). The presence of dust grains is common in these and other
astrophysical and cosmic objects, such as, molecular and interstellar
clouds, planetary and stellar atmospheres, planetary rings, and cometary
tails (Spitzer, Jr. 1978; Goertz 1989; Havnes et al. 1996; Wardle \& Ng
1999; Rotundi et al. 2000). Dust is contained in various plasmas of both
research and technological importance \noindent \noindent (Merlino \& Goree
2004). In space, dust grains vary widely in sizes, from submicrons up to 1
cm and more. Dust particles in both space and the laboratory acquire an
electric charge due to cosmic-ray, radioactive or thermal ionization of the
ambient gas, photoionization, absorption of electrons and ions from the
background plasma, the presence of electron and ion currents, thermoionic
emission, and secondary ionization (Hayashi 1981; Whipple 1981; Meyer-Vernet
1982; Chow et al. 1993; Barkan et al. 1994; Melzer et al. 1994; Mendis \&
Rosenberg 1994; Thomas et al. 1994; Hor\'{a}nyi 1996; Fisher \& Duerbeck
1998; Fortov et al. 1998). The polarity of charged grains in space and the
laboratory can be negative or positive (Meyer-Vernet 1982; Chow et al. 1993;
Mendis \& Rosenberg 1994; Hor\'{a}nyi 1996; Hor\'{a}nyi \& Goertz 1990;
Havnes et al. 2001; Ellis \& Neff 1991). The magnitude of grain charges in
the laboratory can be very large (up to a few orders of magnitude of the
electron charge). The polarity of charged grains depends on their size;
large grains are usually negative and small grains positive (e.g., Chow et
al. 1993; Mendis \& Rosenberg 1994).

The dynamics of dust grains in protostellar and protoplanetary accretion
disks determines the physical processes leading to the formation of planets.
It is generally accepted that small solids (1 cm - 1 m) can be formed due to
chaotic motion and sticking of submicron- and micron-sized dust grains
strongly collision coupled to the surrounding gas. These small solids in
turn can form the kilometer-sized bodies (planetesimals), believed to be the
building blocks for planet formation.

Various physical mechanisms have been discussed for the formation of
planetesimals, such as, collisional coagulation (Adachi et al. 1976; Cuzzi
et al. 1993; Weidenschilling 1995), gravitational instability (Safronov
1969; Goldreich \& Ward 1973; Yamoto \& Sekiya 2004), and the trapping of
particles by large-scale persistent vortices (Barge \& Sommeria 1995;
Lovelace et al. 1999; Klahr \& Bodenheimer 2003; Johansen et al. 2004;
Petersen et al. 2007).

Recently the streaming instability arising from the difference between
velocities of dust grains and gas has been suggested as a possible source
contributing to planetesimal formation (Youdin \& Goodman 2005). These
authors have considered the hydrodynamic instability of the interpenetrating
streams of dust grains and gas coupling via drag forces in the Keplerian
disk. The particle density perturbations generated by this instability could
seed planetesimal formation without self-gravity.

In the papers cited above, which deal with the dynamics of dust grains in
accretion disks in connection with the problem of dust coagulation and
planet formation, hydrodynamic models were applied in which the dust grains
were considered as a neutral fluid or ensemble of individual particles
interacting with the surrounding neutral gas. Thus, electrodynamic processes
were not involved. However, dust grains are generally charged (see
references cited above). In the surface layers and near the star, the
protostellar and protoplanetary accretion disks are multicomponent,
containing electrons, ions, charged grains, and neutral gas. In the
sufficiently dense inner regions, the disk matter likely contains only
charged dust grains and gas (e.g., Wardle \& Ng 1999). Dust grains can be of
various forms. It is clear that electromagnetic forces could play an
important role in dust coagulation and in collective processes within the
disk involving the dust grains. The electromagnetic turbulence favors the
coalescence of charged dust grains. The collisional coagulation due to
electrostatic interaction of dust grains can lead to the formation of
fractal aggregate structures (Matthews et al. 2007). These structures could
build larger-sized solid particles in the process of planet formation.

The necessity of taking into account electromagnetic phenomena in the disk
dynamics is confirmed by observations, which show that astrophysical disks
are turbulent (Hartmann 2000; Carr et al. 2004; Hersant et al. 2005). It is
also known that accretion disks are threaded by the magnetic field
(Hutawarakorn \& Cohen 1999 and 2005; Donati et al. 2005). In the latter, a
direct detection of the magnetic field in the protostellar accretion disk FU
Orionis, including the innermost and densest parts of the disk, is reported.
The surface magnetic field was observed to reach strengths of about 1 kG
close to the center of the disk and several hundred gauss in its innermost
regions. The magnetic field is grossly axisymmetric and has both vertical
and significant azimuthal components. The results of Donati et al. also
suggest that magnetic fields in accretion disks could trigger turbulent
instabilities that would produce enhanced radial accretion and drifts of
ionized plasma through transverse field lines.

The electromagnetic dynamics of accretion disks is determined by physical
parameters, such as, the magnetic field, density, temperature, the
ionization degree, composition, and mass and size of charged dust grains,
all of which vary significantly through the whole disk. So these would
determine the disk dynamics on a local scale. The protostellar and
protoplanetary accretion disks as well as cold molecular and interstellar
clouds are weakly ionized (Wardle \& Ng 1999; Norman \& Heyvaerts 1985; Jin
1996; Bergin et al. 1999; Balbus \& Terquem 2001; Desch 2004; Tscharnuter \&
Gail 2007). Dust grains are, possibly, the primary charge carriers in very
dense, cool nebulae (Wardle \& Ng 1999; Blaes \& Balbus 1994).

One of the sources of electromagnetic turbulence in accretion disks may be
electromagnetic streaming instabilities emerging due to the different
velocities of species in equilibrium. We have studied this issue (Nekrasov
2007) for cold magnetized disks. The dust grains were considered to be
magnetized and strong collisional coupling of neutrals with one of the
charged species could be included. The thermal pressure was not taken into
account. New compressible instabilities were found, with growth rates much
larger than the Keplerian frequency. In a subsequent paper (Nekrasov 2008),
the general theory for electromagnetic streaming instabilities in
multicomponent weakly ionized regions of accretion disks embedded in the
magnetic field has been developed. The compressibility, anisotropic thermal
pressure, and collisions of charged species with neutrals were taken into
account. The equilibrium state was found for the case in which the ions do
not influence the motion of neutrals. However, the neutrals are assumed to
have weak or strong collisional coupling with dust grains. In the perturbed
state, the neutrals have been considered as immobile when the perturbation
frequency is much larger than the collisional frequency of neutrals with
charged species. A dispersion relation has been derived in the most general
form for an arbitrary direction of the perturbation wave vector and for
arbitrary strength of thermal effects. In particular, we found solutions of
the dispersion relation for a case appropriate to definite regions of the
protostellar and protoplanetary disks, where the electrons can be considered
as thermal and magnetized, and ions and dust grains as cold and
unmagnetized. These solutions describe new instabilities of the weakly
ionized disks due to collisions and differences between the stationary
velocities of different charged species.

In the present paper, we study the electromagnetic streaming instabilities\
in the sufficiently ionized regions of the protostellar and protoplanetary
magnetized accretion disks, where the neutrals have strong collisional
coupling simultaneously with ions and dust grains both in the equilibrium
and in the perturbed state. Dust grains and light charged components are
respectively unmagnetized and magnetized, i.e., their cyclotron frequencies
are respectively much smaller or much larger than their orbiting
frequencies. The dust grains are treated as monosized with a constant
charge. For the collisional regime under consideration, we find the
azimuthal and radial stationary velocities of species. We consider the
effect of the change in collisional frequencies due to density perturbations
of species. Taking into account anisotropic thermal pressure, we derive the
general expression for the perturbed velocity of any species that contains
also the perturbed velocity of other species due to collisions. In the cold
limit for horizontally elongated perturbations, we obtain the general
solutions for the perturbed velocities of ions and dust grains incorporating
their mutual influence on each other via collisions with neutrals. Further,
we consider fast processes in which the electromagnetic dynamics of dust
grains is determined by collisions with neutrals but not by their own
motional frequencies. Finally, we investigate unstable perturbations due to
different azimuthal velocities of ions and electrons.

The paper is organized as follows. In \S\ 2 the basic equations are given.
In \S\ 3 we discuss in detail the collisional equilibrium state. The general
solutions for the perturbed velocities and densities of species including
thermal effects are obtained in \S\ 4. The perturbed velocities of species
for vertical perturbations in the cold limit are considered in \S\ 5.
Solutions for the perturbed velocities of species when the dynamics of dust
grains is determined by collisions with neutrals are given in \S\ 6. In \S\
7 we calculate the perturbed electric current. The dispersion relation in
the general case is derived in \S\ 8. The dispersion relation in the case of
strong collisional coupling of dust grains with ions through collisions with
neutrals is considered in \S\ 9. Its unstable solutions are found in \S\ 10.
Discussion of the obtained results is given in \S\ 11. The main points of
the paper are summarized in \S\ 12.

\section{Fundamentals}

\smallskip The fundamental equations in the inertial (nonrotating) reference
frame are the following:
\begin{equation}
\frac{\partial \mathbf{v}_{j}}{\partial t}+\mathbf{v}_{j}\cdot \mathbf{%
\nabla v}_{j}=-\mathbf{\nabla }U-\frac{\mathbf{\nabla }\cdot \mathbf{P}_{j}}{%
m_{j}n_{j}}+\frac{q_{j}}{m_{j}}\mathbf{E+}\frac{q_{j}}{m_{j}c}\mathbf{v}%
_{j}\times \mathbf{B-}\nu _{jn}\left( \mathbf{v}_{j}-\mathbf{v}_{n}\right) ,
\end{equation}
\begin{equation}
\frac{\partial \mathbf{v}_{n}}{\partial t}+\mathbf{v}_{n}\cdot \mathbf{%
\nabla v}_{n}=-\mathbf{\nabla }U-\frac{\mathbf{\nabla }P_{n}}{m_{n}n_{n}}%
-\sum_{j}\nu _{nj}\left( \mathbf{v}_{n}-\mathbf{v}_{j}\right) ,
\end{equation}

\noindent the momentum equations,
\begin{equation}
\frac{\partial n_{j,n}}{\partial t}+\mathbf{\nabla \cdot }n_{j,n}\mathbf{v}%
_{j,n}=0,
\end{equation}

\noindent the continuity equation,
\begin{equation}
\mathbf{\nabla \times E=-}\frac{1}{c}\frac{\partial \mathbf{B}}{\partial t},
\end{equation}
\begin{equation}
\mathbf{\nabla \times B=}\frac{4\pi }{c}\mathbf{j,}
\end{equation}

\noindent and Maxwell's equations, where $\mathbf{j=}\sum_{j}q_{j}n_{j}%
\mathbf{v}_{j}$. Here the index $j=e,i,d$ denotes the electrons, ions, and
dust grains, respectively, and the index $n$ denotes the neutrals. In Eqs.
(1)-(5) $q_{j}$ and $m_{j,n}$ are the charge and mass of species $j$ and
neutrals, $\mathbf{v}_{j,n}$ is the hydrodynamic velocity, $n_{j,n}$ is the
number density, $\mathbf{P}_{j}$ and $P_{n}$ are the thermal pressure tensor
of charged species $j$ and the thermal pressure of neutrals, respectively, $%
\nu _{jn}$ $=\gamma _{jn}m_{n}n_{n}$ ($\nu _{nj}=\gamma _{jn}m_{j}n_{j}$) is
the collisional frequency of charged species (neutrals) with neutrals
(charged species), where $\gamma _{jn}=<\sigma v>_{jn}/(m_{j}+m_{n})$ ($%
<\sigma v>_{jn}$is the rate coefficient for momentum transfer), $U=-GM/R$\
is the gravitational potential of the central object having mass $M$, $%
R=(r^{2}+z^{2})^{1/2}$, $G$ is the gravitational constant, $\mathbf{E}$%
\textbf{\ }and $\mathbf{B}$ are the electric and magnetic fields, and $c$ is
the speed of light in vacuum. We consider wave processes with typical time
scales much larger than the time the light spends to cover the wavelength of
perturbations. In this case one can neglect the displacement current in Eq.
(5), what results in quasineutrality both in electromagnetic and purely
electrostatic perturbations. We use the cylindrical coordinate system $%
(r,\theta ,z),$ where $r$ is the distance from the symmetry axis
and $z$ is the coordinate from the midplane of the disk. The
magnetic field $\mathbf{B}$ includes the external magnetic field
$\mathbf{B}_{0ext}$ of the central object and/or interstellar
medium, the magnetic field $\mathbf{B}_{0cur}$ of the background
current, and the perturbed magnetic field. The pressure
tensor $\mathbf{P}_{j}$ has the form $\mathbf{P}_{j}\mathbf{=}p_{\perp j}%
\mathbf{I+}(p_{\mid \mid j}-p_{\perp j})\mathbf{bb,}$ where $p_{\perp j}$
and $p_{\mid \mid j}$ are the pressure across ($\perp $) and along ($\mid
\mid $) the magnetic field $\mathbf{B}$, $\mathbf{I}$ is the unit tensor,
and $\mathbf{b}$ is the unit vector along $\mathbf{B.}$

Equations (1) and (2) take into account only the collisions between charged
species and neutrals. In the weakly ionized objects these collisions are
dominant. The system of Eqs. (1)-(5) can be applied in the collisionless
regime, if we put $\nu _{jn}=\nu _{nj}=0$, as well as in the regimes of weak
and strong collisional coupling of neutrals with charged components. In Eqs.
(1)-(3) we do not take into account ionization and recombination processes.
We comment on this point in \S\ 11. Self-gravity is not included in the
present paper.

\section{Equilibrium}

We suppose that $\mathbf{B}_{0ext}$ is axisymmetric and
$B_{0\theta ext}=0.$ Such a configuration is typical of the
magnetic field of the central star and/or of disks threaded by the
vertical interstellar magnetic field. We neglect the radial
component of the background magnetic field, considering the
regions of the disk, where vertical components of both the
external magnetic field and the magnetic field induced by the
stationary azimuthal current are dominant.

Let us consider axisymmetric stationary flows of species $j.$ We suppose
that the vertical stationary velocity $v_{j0z}$ is equal to zero. Then the $%
r $ and $\theta $ components of Eq. (1) in the equilibrium take
the form
\begin{equation}
\frac{\partial v_{j0r}^{2}}{2\partial r}-\frac{v_{j0\theta }^{2}}{r}=\frac{%
F_{j0r}}{m_{j}}+\omega _{cj}v_{j0\theta }-\nu _{jn}^{0}(v_{j0r}-v_{n0r}),
\end{equation}
\begin{equation}
\left( \omega _{cj}+\frac{1}{r}\frac{\partial rv_{j0\theta }}{\partial r}%
\right) v_{j0r}=-\nu _{jn}^{0}(v_{j0\theta }-v_{n0\theta }).
\end{equation}

\noindent Here $\omega _{cj}=q_{j}B_{0}/m_{j}c$ is the cyclotron
frequency of species $j$ $(B_{0}=B_{0zext}+B_{0zcur}),$ $\nu
_{jn}^{0}$ is the collisional frequency in the equilibrium, and
\[
F_{j0r}=-m_{j}\frac{\partial U}{\partial r}-\frac{1}{n_{j0}}\frac{\partial
p_{\perp 0j}}{\partial r}+q_{j}E_{0r},
\]
where $E_{0r}$ is the background electric field.

Equations (6) and (7) determine the stationary velocity $\mathbf{v}_{j0}$ of
different charged components due to the action of the electric (radial),
magnetic (vertical), and gravitational fields as well as the thermal
pressure and collisions with neutrals. Due to the latter effect the radial
velocity emerges [Eq. (7)]. Solving Eqs. (6) and (7), we will consider two
cases in which there are magnetized, $\omega _{cj}v_{j0\theta }\gg
v_{j0r,\theta }^{2}/r$, and unmagnetized, $\omega _{cj}v_{j0\theta }\ll
v_{j0r,\theta }^{2}/r$, charged species. The condition $\omega _{cj}\gg (\ll
)v_{j0\theta }/r$ denotes that the cyclotron frequency is much larger
(smaller) than the orbital frequency [we use below, as usual, the term
''magnetized'' (''unmagnetized'') also in the case in which $\omega _{cj}\gg
(\ll )\nu _{jn}^{0}$]. \smallskip

For magnetized species we have the following solutions of Eqs. (6) and (7):
\begin{equation}
v_{j0\theta }=-\frac{\omega _{cj}}{\omega _{cj}^{2}+\nu _{jn}^{2}}\frac{%
F_{j0r}}{m_{j}}+\frac{\nu _{jn}^{2}}{\omega _{cj}^{2}+\nu _{jn}^{2}}%
v_{n0\theta }-\frac{\omega _{cj}\nu _{jn}}{\omega _{cj}^{2}+\nu _{jn}^{2}}%
v_{n0r},
\end{equation}
\begin{equation}
v_{j0r}=\frac{\nu _{jn}}{\omega _{cj}^{2}+\nu _{jn}^{2}}\frac{F_{j0r}}{m_{j}}%
+\frac{\omega _{cj}\nu _{jn}}{\omega _{cj}^{2}+\nu _{jn}^{2}}v_{n0\theta }+%
\frac{\nu _{jn}^{2}}{\omega _{cj}^{2}+\nu _{jn}^{2}}v_{n0r}.
\end{equation}

\noindent Here, for simplicity, the index $0$ at $\nu _{jn}^{0}$ has been
omitted.

We will consider the case in which simultaneous collisions of neutrals with
two charged components are important. We suppose that one of these
components is the heavy unmagnetized dust grains $d$ and another component,
which will be denoted by the index $j$, represents the ions or the light
magnetized dust grains (the latter case is appropriate for systems
containing only two dust components). Due to the small electron mass the
collisions of neutrals with electrons are neglected. Below in this section,
the corresponding condition is given. Then, the equation of motion (2) for
neutrals takes the form
\begin{equation}
\frac{\partial v_{n0r}^{2}}{2\partial r}-\frac{v_{n0\theta }^{2}}{r}=\frac{%
F_{n0r}}{m_{n}}-\nu _{nj}(v_{n0r}-v_{j0r})-\nu _{nd}(v_{n0r}-v_{d0r}),
\end{equation}
\begin{equation}
\frac{1}{r}\frac{\partial rv_{n0\theta }}{\partial r}v_{n0r}=-\nu
_{nj}(v_{n0\theta }-v_{j0\theta })-\nu _{nd}(v_{n0\theta }-v_{d0\theta }).
\end{equation}

\noindent Here $\nu _{nj,d}=\nu _{nj,d}^{0}$. For unmagnetized dust grains
Eqs. (6) and (7) have the form
\begin{equation}
\frac{\partial v_{d0r}^{2}}{2\partial r}-\frac{v_{d0\theta }^{2}}{r}=\frac{%
F_{d0r}}{m_{d}}-\nu _{dn}(v_{d0r}-v_{n0r}),
\end{equation}
\begin{equation}
\frac{1}{r}\frac{\partial rv_{d0\theta }}{\partial r}v_{d0r}=-\nu
_{dn}(v_{d0\theta }-v_{n0\theta }).
\end{equation}

To solve the system (10)-(13) we adopt, for simplicity, that $v_{n,d0\theta
}\sim r^{-\alpha _{1}}$ and $v_{n,d0r}\sim r^{-\alpha _{2}}.$ Then we can
substitute the derivatives in Eqs. (10)-(13) by the algebraic terms. We find
further the expressions for $v_{n0r}-v_{j0r}$ and $v_{n0\theta }-v_{j0\theta
}$, using Eqs. (8) and (9), and substitute them in Eqs. (10) and (11). From
Eqs. (11) and (13) we can find the radial velocities $v_{n0r}$ and $v_{d0r}$%
:
\begin{equation}
v_{n0r}=-\frac{r}{a_{j}}\frac{\nu _{nj}^{*}\omega _{cj}}{\omega
_{cj}^{2}+\nu _{jn}^{2}}\left( \frac{F_{j0r}}{m_{j}v_{n0\theta }}+\omega
_{cj}\right) -\frac{\nu _{nd}^{*}r}{a_{j}}\left( 1-x\right) ,
\end{equation}
\begin{equation}
v_{d0r}=-\nu _{dn}^{*}r(1-\frac{1}{x}),
\end{equation}

\noindent where $x=v_{d0\theta }/v_{n0\theta }$ and the index $*$ denotes
the modified collisional frequencies $\nu _{nj,d}^{*}=\nu _{nj,d}/(1-\alpha
_{1})$ and $\nu _{dn}^{*}=\nu _{dn}/(1-\alpha _{1})$. The parameter $a_{j}$
in Eq. (14) is
\begin{equation}
a_{j}=1+b_{j}=1+\frac{r}{v_{n0\theta }}\frac{\nu _{nj}^{*}\nu _{jn}\omega
_{cj}}{\omega _{cj}^{2}+\nu _{jn}^{2}}.
\end{equation}

\noindent After substitution of expressions (14) and (15) into Eqs. (10) and
(12) we obtain two equations containing only the azimuthal velocities of
neutrals and dust grains (except the terms proportional to $\alpha _{2}$):
\[
-\frac{v_{n0\theta }^{2}}{r}=\alpha _{2}\frac{v_{n0r}^{2}}{r}+\frac{F_{n0r}}{%
m_{n}}+\frac{\nu _{nj}}{\omega _{cj}^{2}+\nu _{jn}^{2}}\left[ \nu
_{jn}+\left( \frac{\nu _{nj}^{*}\omega _{cj}^{2}}{\omega _{cj}^{2}+\nu
_{jn}^{2}}+\nu _{nd}^{*}\right) \frac{\omega _{cj}r}{a_{j}v_{n0\theta }}%
\right] \left( \frac{F_{j0r}}{m_{j}}+\omega _{cj}v_{n0\theta }\right)
\]
\begin{equation}
+\nu _{nd}r\left( 1-x\right) \left[ \frac{\nu _{dn}^{*}}{x}+\frac{1}{a_{j}}%
\left( \frac{\nu _{nj}^{*}\omega _{cj}^{2}}{\omega _{cj}^{2}+\nu _{jn}^{2}}%
+\nu _{nd}^{*}\right) \right] ,
\end{equation}
\begin{equation}
-\frac{v_{d0\theta }^{2}}{r}=\alpha _{2}\frac{v_{d0r}^{2}}{r}+\frac{F_{d0r}}{%
m_{d}}-\frac{r}{a_{j}v_{n0\theta }}\frac{\nu _{dn}\nu _{nj}^{*}\omega _{cj}}{%
\omega _{cj}^{2}+\nu _{jn}^{2}}\left( \frac{F_{j0r}}{m_{j}}+\omega
_{cj}v_{n0\theta }\right) -\nu _{dn}r\left( 1-x\right) \left( \frac{\nu
_{dn}^{*}}{x}+\frac{\nu _{nd}^{*}}{a_{j}}\right) .
\end{equation}

\noindent For simplicity, we have retained in these equations the radial
velocities in terms proportional to $\alpha _{2}$. Multiplying Eq. (17) by $%
x^{2}$ and equating it to Eq. (18), we obtain the following equation:
\[
x^{2}\frac{F_{n0r}}{m_{n}}-\frac{F_{d0r}}{m_{d}}+\frac{\nu _{nj}}{\omega
_{cj}^{2}+\nu _{jn}^{2}}\left( \frac{F_{j0r}}{m_{j}}+\omega _{cj}v_{n0\theta
}\right) \left[ x^{2}\nu _{jn}+\left( x^{2}\frac{\nu _{nj}^{*}\omega
_{cj}^{2}}{\omega _{cj}^{2}+\nu _{jn}^{2}}+x^{2}\nu _{nd}^{*}+\nu
_{dn}^{*}\right) \frac{\omega _{cj}r}{a_{j}v_{n0\theta }}\right]
\]
\begin{equation}
+r\left( 1-x\right) \left[ x\nu _{nd}\nu _{dn}^{*}+\frac{\nu _{nd}^{*}\nu
_{dn}}{a_{j}}+\frac{\nu _{dn}\nu _{dn}^{*}}{x}+x^{2}\frac{\nu _{nd}}{a_{j}}%
\left( \frac{\nu _{nj}^{*}\omega _{cj}^{2}}{\omega _{cj}^{2}+\nu _{jn}^{2}}%
+\nu _{nd}^{*}\right) \right] -\alpha _{2}\frac{v_{d0r}^{2}}{r}+x^{2}\alpha
_{2}\frac{v_{n0r}^{2}}{r}=0.
\end{equation}

In the paper (Nekrasov 2008), we have considered the case of the
weak collisional coupling of neutrals with ions in the steady
state, which is appropriate for weakly ionized disks. The
parameter $b_{j}$ [see Eq. (16)]
for ions has been supposed to be much less than unity, in which case $%
a_{i}\simeq 1$. This condition can be satisfied for magnetized, $\omega
_{ci}\gg \nu _{in}$, or unmagnetized, $\omega _{ci}\ll \nu _{in}$, ions, if
the orbital (Keplerian) frequency of neutrals $v_{n0\theta }/r$ is less than
the collisional frequency of neutrals with ions $\nu _{ni}$. In the case $%
\omega _{ci}\sim \nu _{in}$ it should be $\nu _{ni}\ll v_{n0\theta }/r$. In
the present paper, we will consider the opposite case in which $b_{j}\gg 1$,
i.e.,
\begin{equation}
\frac{\nu _{jn}\omega _{cj}}{\omega _{cj}^{2}+\nu _{jn}^{2}}\nu _{nj}^{*}\gg
\frac{v_{n0\theta }}{r}.
\end{equation}

\noindent Thus, we explore here the regions of the disk, where the medium is
sufficiently ionized and where a strong collisional coupling between
neutrals and species $j$ (ions or light dust grains) takes place.

We can find the analytical solution of Eq. (19) in the case in which the
azimuthal velocities of neutrals and dust grains are close to one another,
i.e.,
\[
x=1+\delta ,\delta \ll 1.
\]

\noindent This condition is easily satisfied in disks, where the collisional
coupling of dust grains and neutrals is strong (see below). We neglect
further the contribution of terms proportional to $\delta $ in the first and
third terms on the left-hand side of Eq. (19) as well as the contribution of
two last terms in this equation. The corresponding conditions, taking into
account the inequality (20), have the form
\[
\nu ^{2}\gg \frac{2}{r}\left[ \frac{F_{n0r}}{m_{n}}+\frac{\nu _{nj}+\nu _{nd}%
}{\nu _{jn}}\left( \frac{F_{j0r}}{m_{j}}+\omega _{cj}v_{n0\theta }\right)
\right] ;2\alpha _{2}\frac{v_{n0r}^{2}}{r^{2}},
\]
\begin{equation}
\delta \nu ^{2}\gg \alpha _{2}\frac{1}{r^{2}}\left(
v_{n0r}^{2}-v_{d0r}^{2}\right) ,
\end{equation}

\noindent where $\nu ^{2}\simeq \nu _{dn}^{*}\left( \nu _{dn}+\nu
_{nd}\right) $. We suppose that $b_{j}\nu _{dn}\gg \nu _{nd}$. This
condition is, obviously, satisfied. Conditions (21) imply strong
dust-neutrals collisional coupling. Then we obtain from Eq. (19) the
following solution for $\delta $:
\begin{equation}
\delta r\nu ^{2}=\frac{F_{n0r}}{m_{n}}-\frac{F_{d0r}}{m_{d}}+\frac{1}{\nu
_{jn}}\left( \nu _{nj}+\nu _{nd}+\nu _{dn}\right) \left( \frac{F_{j0r}}{m_{j}%
}+\omega _{cj}v_{n0\theta }\right) .
\end{equation}

The second equation connecting $\delta $ and $v_{n0\theta }$ is obtained
from Eq. (18) under conditions (20) and
\begin{equation}
\delta \nu _{dn}\nu _{dn}^{*}\gg \frac{v_{d0\theta }^{2}}{r^{2}}+\alpha _{2}%
\frac{v_{d0r}^{2}}{r^{2}}.
\end{equation}

\noindent This inequality also implies strong collisional coupling of dust
grains with neutrals. Then we have
\begin{equation}
\delta r\nu _{dn}\nu _{dn}^{*}=-\frac{F_{d0r}}{m_{d}}+\frac{\nu _{dn}}{\nu
_{jn}}\left( \frac{F_{j0r}}{m_{j}}+\omega _{cj}v_{n0\theta }\right) .
\end{equation}

We can find from Eqs. (22) and (24) the expressions for $v_{n0\theta }$ and $%
\delta $. The velocity $v_{n0\theta }$ is equal to
\begin{equation}
-\omega _{cj}v_{n0\theta }=\frac{F_{j0r}}{m_{j}}+\frac{\nu _{jn}}{\nu _{nj}}%
\left( \frac{F_{n0r}}{m_{n}}+\frac{\nu _{nd}}{\nu _{dn}}\frac{F_{d0r}}{m_{d}}%
\right) .
\end{equation}

\noindent It can be shown that the condition
\[
b_{j}\gg \frac{\omega _{cj}^{2}}{\omega _{cj}^{2}+\nu _{jn}^{2}}\frac{\nu
_{nj}}{\nu _{dn}}
\]

\noindent must be satisfied for the solution (25) to have the present form.
It is obvious that this inequality can easily be realized, if $b_{j}\nu
_{dn}\gg \nu _{nj,d}$. The value $\delta $ we find from Eqs. (24) and (25),
\begin{equation}
\delta r\nu _{dn}^{*}=-\frac{1}{\nu _{nj}}\left( \frac{F_{n0r}}{m_{n}}+\frac{%
\nu _{nj}+\nu _{nd}}{\nu _{dn}}\frac{F_{d0r}}{m_{d}}\right) .
\end{equation}

The azimuthal velocity of dust grains is equal to
\begin{equation}
v_{d0\theta }=v_{n0\theta }(1+\delta ).
\end{equation}

\noindent The radial velocities of neutrals and dust grains are found from
Eqs. (12) [or (14)] and (15):
\[
v_{n0r}=-\frac{F_{d0r}}{\nu _{dn}m_{d}}-\delta r\nu _{dn}^{*},
\]
\begin{equation}
v_{d0r}=-\delta r\nu _{dn}^{*}.
\end{equation}

Let us now consider the velocities $v_{j0\theta ,r}.$ Substituting the
velocities $v_{n0\theta ,r}$ found above into Eqs. (8) and (9), we obtain $%
v_{j0\theta }\simeq v_{n0\theta }$ and $v_{j0r}\ll v_{j0\theta },v_{n,d0r}$.

The electron velocities under condition $\omega _{ce}^{2}\gg \nu _{en}^{2}$
are the following [see Eqs. (8) and (9)]:
\[
v_{e0\theta }=-\frac{F_{e0r}}{m_{e}\omega _{ce}}+\frac{\nu _{en}^{2}}{\omega
_{ce}^{2}}v_{n0\theta }-\frac{\nu _{en}}{\omega _{ce}}v_{n0r},
\]
\begin{equation}
v_{e0r}=\frac{\nu _{en}}{\omega _{ce}^{2}}\frac{F_{e0r}}{m_{e}}+\frac{\nu
_{en}}{\omega _{ce}}v_{n0\theta }+\frac{\nu _{en}^{2}}{\omega _{ce}^{2}}%
v_{n0r}.
\end{equation}

We do not take into account the collisions of neutrals with electrons. It
can be shown that the corresponding condition has the form
\[
\nu _{ne}\ll \nu _{nj}\frac{\omega _{cj}^{2}}{\omega _{cj}^{2}+\nu _{jn}^{2}}%
.
\]

\noindent We suppose that this inequality is satisfied.

\section{Perturbed velocities and densities of species:
General solutions}

In the present paper, we do not treat perturbations connected with the
background pressure gradients. Thus, we exclude from our consideration the
drift and internal gravity waves. We take into account the induced reaction
of neutrals on the perturbed motion of charged species. This effect is
important, if the ionization degree of medium is sufficiently high. We also
include the effect of perturbation of the collisional frequencies due to
density perturbations of charged species and neutrals. This effect emerges
when there are different background velocities of species, as occurs in
accretion disks. Then the momentum equations (1) and (2) in the linear
approximation take the form
\begin{equation}
\frac{\partial \mathbf{v}_{j1}}{\partial t}+\mathbf{v}_{j0}\cdot \mathbf{%
\nabla v}_{j1}+\mathbf{v}_{j1}\cdot \mathbf{\nabla v}_{j0}=-\frac{\mathbf{P}%
_{0j}}{m_{j}n_{j0}^{2}}\mathbf{\nabla }n_{j1}+\frac{q_{j}}{m_{j}}\mathbf{E}%
_{1}+\frac{q_{j}}{m_{j}c}\mathbf{v}_{j0}\times \mathbf{B}_{1}+\frac{q_{j}}{%
m_{j}c}\mathbf{v}_{j1}\times \mathbf{B}_{0}+\mathbf{C}_{j1},
\end{equation}

\begin{equation}
\frac{\partial \mathbf{v}_{n1}}{\partial t}+\mathbf{v}_{n0}\cdot \mathbf{%
\nabla v}_{n1}+\mathbf{v}_{n1}\cdot \mathbf{\nabla v}_{n0}=-v_{Tn}^{2}\frac{%
\mathbf{\nabla }n_{n1}}{n_{n0}}\mathbf{+C}_{n1},
\end{equation}

\noindent In these equations, the collisional terms $\mathbf{C}_{j1}$ and $%
\mathbf{C}_{n1}$ are equal
\[
\mathbf{C}_{j1}=-\nu _{jn}^{0}\mathbf{v}_{j1}+\mathbf{R}_{j1},
\]
\[
\mathbf{C}_{n1}=-\nu _{n}^{0}\mathbf{v}_{n1}+\mathbf{R}_{n1},
\]

\noindent where $\nu _{jn}^{0}=\gamma _{jn}m_{n}n_{n0}$, $\nu
_{n}^{0}=\sum_{j}\nu _{nj}^{0}$, $\nu _{nj}^{0}=\gamma _{nj}m_{j}n_{j0}$.
The index $1$ denotes the quantities of the first order of magnitude. Here $%
\mathbf{P}_{0j}=\gamma _{\perp j}p_{\perp 0j}\mathbf{I+(}\gamma _{\mid \mid
j}p_{\mid \mid 0j}-\gamma _{\perp j}p_{\perp 0j})\mathbf{bb},$ $\gamma
_{\perp j}=2,$ $\gamma _{\mid \mid j}=3$ for magnetized particles $\omega
_{cj}\gg \nu _{jn}.$ For charged species we use here the double adiabatic
dependence of temperature perturbations on the density perturbations
analogous to the double adiabatic law of Chew et al. (1956) for the
collisionless plasma. For unmagnetized charged particles, $\omega _{cj}\ll
\nu _{jn},$ we have $\mathbf{P}_{0j}=\gamma _{j}p_{0j}\mathbf{I,}$ $\gamma
_{j}=5/3.$ In the case of Boltzmann's distribution of the number density due
to movement of particles along the background magnetic field we have $\gamma
_{\mid \mid j}=1.$ The neutrals are also considered as adiabatic. Then $%
\mathbf{\nabla }P_{n1}/m_{n}n_{n0}=v_{Tn}^{2}\mathbf{\nabla }n_{n1}/n_{n0}$,
where $v_{Tn}^{2}=\gamma _{n}T_{n0}/m_{n}$, $T_{n0}$ is the unperturbed
temperature of neutrals.

The term $\mathbf{R}_{j1},$
\[
\mathbf{R}_{j1}=\nu _{jn}^{0}\left[ \mathbf{v}_{n1}+\frac{n_{n1}}{n_{n0}}%
\left( \mathbf{v}_{n0}-\mathbf{v}_{j0}\right) \right] ,
\]

\noindent describes the back reaction of charged components due to influence
of the perturbed motion of neutrals and the term $\mathbf{R}_{n1},$%
\[
\mathbf{R}_{n1}=\sum_{j}\nu _{nj}^{0}\left[ \mathbf{v}_{j1}+\frac{n_{j1}}{%
n_{j0}}\left( \mathbf{v}_{j0}-\mathbf{v}_{n0}\right) \right] ,
\]

\noindent leads to the induced perturbation of the velocity of neutrals. The
collisional terms proportional to $n_{n1}$ and $n_{j1}$ take into account
the dependence of the collisional frequencies $\nu _{jn}$ and $\nu _{nj}$ on
the density perturbations of species.

In this section, we will consider the general case in which the wave vector $%
\mathbf{k=(}k_{r},m\theta ,k_{z})$ of perturbations has three components. We
provide the normal mode analysis, assume the perturbations to be
proportional to $\exp (ik_{r}r+im\theta +ik_{z}z-i\omega t),$ and use the
local approximation $|\mathbf{k}|r\gg 1.$ Then we find the Fourier
amplitudes of the density perturbations from the linearized version of the
continuity equation (3)
\[
\frac{n_{j,n1k}}{n_{j,n0}}=\frac{\mathbf{kv}_{j,n1k}}{\omega _{j,n}},
\]
where $\omega _{j,n}=\omega -\mathbf{kv}_{j,n0},$ $k=\{\mathbf{k,}\omega \}.$
Substituting $n_{n1}$ into Eq. (31) and solving this equation in the Fourier
representation, we will find the velocity $\mathbf{v}_{n1k},$ whose
components and the expression for $n_{n1k}$ are given in the Appendix A.

In the protostellar and protoplanetary disks, the collisions of neutrals
with charged species are frequent enough that the condition $\omega _{\nu
n}\gg \Omega _{n}$ is satisfied (for systems with the weak collisions it
should be supposed that $\omega _{n}\gg \Omega _{n}$). Then, using the
equation for $n_{n1k}$ from the Appendix A and the expression for $\mathbf{R}%
_{n1}$, we obtain
\begin{equation}
D_{n}\frac{n_{n1k}}{n_{n0}}=i\omega _{\nu n}\sum_{j}\nu _{nj}^{0}\frac{%
n_{j1k}}{n_{j0}}.
\end{equation}

In the present paper, we will further neglect the thermal effects of species
when considering some specific cases. For neutrals, the necessary condition
for this cold limit has the form $\delta _{lmn}\ll 1$, where $l,m=r,\theta
,z $. In this case, the perturbed velocity and density of neutrals given in
the Appendix A and Eq. (32), correspondingly, will be equal to
\begin{equation}
\mathbf{v}_{n1k}=\frac{i}{\omega _{\nu n}}\mathbf{R}_{n1k}
\end{equation}

\noindent and
\begin{equation}
\frac{n_{n1k}}{n_{n0}}=\frac{i}{\omega _{\nu n}}\sum_{j}\nu _{nj}^{0}\frac{%
n_{j1k}}{n_{j0}}.
\end{equation}

\noindent Let us substitute the expressions (33) and (34) into $\mathbf{R}%
_{j1k}.$ Then the collisional term $\mathbf{C}_{j1k}$ can be written in the
form
\begin{equation}
\mathbf{C}_{j1k}=-\nu _{j}\mathbf{v}_{j1k}+\mathbf{Q}_{j1k},
\end{equation}
where
\[
\nu _{j}=\nu _{jn}^{0}\left( 1-i\frac{\nu _{nj}^{0}}{\omega _{\nu n}}\right)
,
\]
\[
\mathbf{Q}_{j1k}=i\frac{\nu _{jn}^{0}}{\omega _{\nu n}}\sum_{l\neq j}\nu
_{nl}^{0}\left[ \mathbf{v}_{l1k}+\frac{n_{l1k}}{n_{l0}}\left( \mathbf{v}%
_{l0}-\mathbf{v}_{j0}\right) \right] .
\]

\noindent Here the index $l$ denotes the charged components. The collisional
term $\mathbf{Q}_{j1}$ describes the reaction of the species $j$ on the
collisions of neutrals with another charged components.

In the weakly ionized plasma the frequency of collisions of neutrals with
charged components can be small compared to the frequency of perturbations,
which depends on the disk parameters, $\omega _{n}\gg \nu _{n}^{0}$. In this
case, the induced dynamics of neutrals can be neglected, i.e., $\nu
_{j}\approx \nu _{jn}^{0}$ and $\mathbf{Q}_{j1k}\approx \mathbf{0}$. Such a
situation was considered by Nekrasov (2008). In the sufficiently ionized
plasma the opposite case can be realized in which $\nu _{n}^{0}\gg \omega
_{n}$. Then the collisional frequency $\nu _{j}$ takes the form
\begin{equation}
\nu _{j}=\frac{\nu _{jn}^{0}}{\nu _{n}^{0}}\left( \sum_{l\neq j}\nu
_{nl}^{0}-i\omega _{n}\frac{\nu _{nj}^{0}}{\nu _{n}^{0}}\right) .
\end{equation}

\noindent We see that when there is only one charged component with which
the neutrals collide, the inertia of neutrals is added to the inertia of
this charged component, i.e., $\nu _{j}=-i\omega _{n}(\nu _{jn}^{0}/\nu
_{nj}^{0})$. At the same time the value $\mathbf{Q}_{j1}=\mathbf{0.}$ Thus,
in other respects the plasma stays formally collisionless. However, the
presence of other charged components with which neutrals collide with the
sufficiently high frequency, $\nu _{nl}^{0}\gg \omega _{n}$, changes the
situation dramatically: The plasma stays collisional and the value $\mathbf{Q%
}_{j1}$ contributes to the momentum equation of species $j$. Such conditions
are realized in the ionized regions of protostellar and protoplanetary
disks. This case will be considered in the present paper.

The solution of Eq. (30) with $\mathbf{C}_{j1k}$ defined by Eq. (35) and the
expression for $n_{j1k}$ are given in the general form in the Appendix B.

We will also neglect the thermal effects for the charged species. The
necessary condition is $\delta _{lmj}\ll 1$, where $l,m=r,\theta ,z$. We
consider further the case in which the neutrals collide simultaneously with
ions and dust grains. As a result, the mutual influence of perturbations of
ions and dust grains via perturbations of neutrals arises. The corresponding
system of equations describing this process is given in the Appendix C.

\section{Perturbed velocities of species in the case $k_{z}\neq 0,$ $%
k_{r}=k_{\theta }=0$: General solutions}

We have obtained the exact solutions of the system of equations for $%
v_{i1r,\theta }$ and $v_{d1r,\theta }$ given in the Appendix C for the
horizontally elongated perturbations when $k_{z}\neq 0,$ $k_{r}=k_{\theta
}=0.$ These solutions and the general solutions for $v_{i,d1z}$ are given in
the Appendix D.

If we set $\mu _{i}=\mu _{d}=0$ in Eqs. (C.1) and (C.2), i.e., if we neglect
the collisions of neutrals with ions and dust grains ($\omega _{n}\gg \nu
_{n}^{0}$), then we obtain $D_{i,d}v_{i,d1r}=-H_{i,d1r}$, $%
D_{i,d}v_{i,d1\theta }=-H_{i,d1\theta }$, and $v_{i,d1z}=(i/\omega _{\nu
i,d})G_{i,d1z}$. In this case, the ions and dust grains move independently
from each other under the action of the electromagnetic forces. However, in
the case $\omega _{n}\ll \nu _{n}^{0}$ the movement of dust grains depends
on the movement of ions, and vice versa (see below).

The general solutions for the perturbed electron velocity in the case $%
k_{z}\neq 0,$ $k_{r}=k_{\theta }=0$ are given in the Appendix E. One can see
that the collisions of neutrals with ions and dust grains influence the
collisions of electrons with neutrals.

\section{Solutions for $\mathbf{v}_{j1k}$ in the case of
neglect the own dynamics of dust grains}

The own frequencies of dust grains, the plasma and cyclotron frequencies,
are very low because of their large mass. Therefore, at the absence of
collisions of species with dust grains the latter can stay immobile in the
electromagnetic perturbations having sufficiently high frequency. However,
for example, through collisions with neutrals, the dust grains can be
involved in the fast electromagnetic perturbations induced by the ions and
electrons when the ions have also strong collisional coupling with neutrals.

Let us neglect in Eq. (D.3) the terms proportional to $\omega _{d1}$ and $%
\omega _{d2}$ under conditions
\begin{equation}
\omega _{\nu d}^{2}\gg \left( 1+\frac{\omega _{\nu i}^{2}}{\omega _{ci}^{2}}%
\right) \omega _{d1}\omega _{d2};\mu _{i}\mu _{d}\frac{\omega _{d1,2}}{%
\omega _{ci}}.
\end{equation}
Note, that for frequencies of perturbations much less than the collisional
frequencies, $\nu _{i,d}\gg \omega _{i,d}$, these conditions can be written
in the form $\nu _{d}\gg (1+\nu _{i}/\omega _{ci})\omega _{d1,2}$. We have
taken into account that $\omega _{ci}\gg \Omega _{i}$. Then the value $A$
will be equal to $A=-\triangle ,$ where $\triangle =\lambda _{\omega
}^{2}+\omega _{\nu d}^{2}\omega _{ci}^{2}.$ In the case (37), one can also
neglect the action of the electromagnetic forces on the dust grains. Then
the expressions (D.1) and (D.2) take the form,
\[
v_{i1r}=\frac{\omega _{\nu d}}{\triangle }\left( \omega _{\nu d}\omega
_{ci}G_{i1\theta }+\lambda _{\omega }G_{i1r}\right) ,
\]
\begin{equation}
v_{i1\theta }=\frac{\omega _{\nu d}}{\triangle }\left( -\omega _{\nu
d}\omega _{ci}G_{i1r}+\lambda _{\omega }G_{i1\theta }\right) .
\end{equation}

\noindent For the dust grains we have
\[
v_{d1r}=-\frac{\mu _{i}}{\omega _{\nu d}}v_{i1r}-\frac{\mu _{i}}{\omega
\lambda _{\omega }}k_{z}w_{r}G_{i1z},
\]
\begin{equation}
v_{d1\theta }=-\frac{\mu _{i}}{\omega _{\nu d}}v_{i1\theta }-\frac{\mu _{i}}{%
\omega \lambda _{\omega }}k_{z}w_{\theta }G_{i1z}.
\end{equation}
The vertical velocities $v_{i1z}$ and $v_{d1z}$ are equal in the case under
consideration,
\[
\lambda _{\omega }v_{i1z}=\omega _{\nu d}G_{i1z},
\]
\begin{equation}
\lambda _{\omega }v_{d1z}=-\mu _{i}G_{i1z}.
\end{equation}

We can find now the perturbed electron velocity. Substituting the velocities
$\mathbf{v}_{i,d1}$ defined by Eqs. (38) and (39) in the expressions for the
components of the electron velocity given in the Appendix E, we obtain
\begin{eqnarray*}
D_{e}v_{e1r} &=&-H_{e1r}+\frac{\eta _{\omega }}{\triangle }\left( \lambda
_{\omega }\omega _{\nu e}-i\omega _{\nu d}\omega _{ce}\omega _{ci}\right)
G_{i1r}+\frac{\eta _{\omega }}{\triangle }\left( i\lambda _{\omega }\omega
_{ce}+\omega _{\nu d}\omega _{\nu e}\omega _{ci}\right) G_{i1\theta } \\
&&+\frac{\eta _{\omega }}{\omega \lambda _{\omega }}\varepsilon
_{er}k_{z}G_{i1z},
\end{eqnarray*}
\begin{eqnarray}
D_{e}v_{e1\theta } &=&-H_{e1\theta }-\frac{\eta _{\omega }}{\triangle }%
\left( i\lambda _{\omega }\omega _{ce}+\omega _{\nu d}\omega _{\nu e}\omega
_{ci}\right) G_{i1r}+\frac{\eta _{\omega }}{\triangle }\left( \lambda
_{\omega }\omega _{\nu e}-i\omega _{\nu d}\omega _{ce}\omega _{ci}\right)
G_{i1\theta }  \nonumber \\
&&+\frac{\eta _{\omega }}{\omega \lambda _{\omega }}\varepsilon _{e\theta
}k_{z}G_{i1z},  \nonumber
\end{eqnarray}
\begin{equation}
v_{e1z}=\frac{i}{\omega _{\nu e}}G_{e1z}+\frac{\eta _{\omega }}{\omega _{\nu
e}\lambda _{\omega }}G_{i1z},
\end{equation}
\noindent where
\begin{eqnarray*}
\eta _{\omega } &=&\mu _{i}\mu _{ed}-\omega _{\nu d}\mu _{ei}, \\
\varepsilon _{er} &=&\omega _{\nu e}\left( v_{i0r}-v_{e0r}\right) +i\omega
_{ce}\left( v_{i0\theta }-v_{e0\theta }\right) , \\
\varepsilon _{e\theta } &=&\omega _{\nu e}\left( v_{i0\theta }-v_{e0\theta
}\right) -i\omega _{ce}\left( v_{i0r}-v_{e0r}\right) .
\end{eqnarray*}

\section{Perturbed electric current: General expressions}

In this section, we find the perturbed electric current
\[
\mathbf{j}_{1}\mathbf{=}\sum_{j}q_{j}n_{j0}\left( \mathbf{v}_{j1}+\frac{k_{z}%
\mathbf{v}_{j0}}{\omega }v_{j1z}\right) .
\]

\noindent Using Eqs. (38)-(41), we can write the components of this current
in the form
\begin{eqnarray*}
j_{1r} &=&\left[ \frac{q_{i}}{m_{i}\triangle }W_{rr}+W_{zrr}\right]
E_{1r}+\left[ \frac{q_{i}}{m_{i}\triangle }W_{r\theta }+W_{zr\theta }\right]
E_{1\theta }+W_{zrz}E_{1z}, \\
j_{1\theta } &=&\left[ \frac{q_{i}}{m_{i}\triangle }W_{\theta r}+W_{z\theta
r}\right] E_{1r}+\left[ \frac{q_{i}}{m_{i}\triangle }W_{\theta \theta
}+W_{z\theta \theta }\right] E_{1\theta }+W_{z\theta z}E_{1z}, \\
j_{1z} &=&W_{zzr}E_{1r}+W_{zz\theta }E_{1\theta }+W_{zzz}E_{1z}.
\end{eqnarray*}

\noindent The expressions for $W_{rr,\theta }$, $W_{\theta r,\theta }$, $%
W_{zrr,\theta ,z}$, $W_{z\theta r,\theta ,z}$, and $W_{zzr,\theta ,z}$ in
the general form are given in the Appendix F.

\section{Dispersion relation in the general form}

From Maxwell's equations (4) and (5) we have,
\begin{equation}
n_{z}^{2}E_{1r,\theta }=\frac{4\pi i}{\omega }j_{1r,\theta },0=j_{1z}.
\end{equation}
Using the components of the perturbed electric current given in \S\ 7, we
will find from the system (42) the dispersion relation
\begin{equation}
\left( n_{z}^{2}\varepsilon _{zz}-\varepsilon _{rr}\varepsilon
_{zz}+\varepsilon _{rz}\varepsilon _{zr}\right) \left( n_{z}^{2}\varepsilon
_{zz}-\varepsilon _{\theta \theta }\varepsilon _{zz}+\varepsilon _{\theta
z}\varepsilon _{z\theta }\right) -\left( -\varepsilon _{r\theta }\varepsilon
_{zz}+\varepsilon _{rz}\varepsilon _{z\theta }\right) \left( -\varepsilon
_{\theta r}\varepsilon _{zz}+\varepsilon _{zr}\varepsilon _{\theta z}\right)
=0.
\end{equation}

\noindent Here
\begin{eqnarray*}
\varepsilon _{rr} &=&\frac{4\pi i}{\omega }\left[ \frac{q_{i}}{%
m_{i}\triangle }W_{rr}+W_{zrr}\right] ,\varepsilon _{r\theta }=\frac{4\pi i}{%
\omega }\left[ \frac{q_{i}}{m_{i}\triangle }W_{r\theta }+W_{zr\theta
}\right] ,\varepsilon _{rz}=\frac{4\pi i}{\omega }W_{zrz}, \\
\varepsilon _{\theta r} &=&\frac{4\pi i}{\omega }\left[ \frac{q_{i}}{%
m_{i}\triangle }W_{\theta r}+W_{z\theta r}\right] ,\varepsilon _{\theta
\theta }=\frac{4\pi i}{\omega }\left[ \frac{q_{i}}{m_{i}\triangle }W_{\theta
\theta }+W_{z\theta \theta }\right] ,\varepsilon _{\theta z}=\frac{4\pi i}{%
\omega }W_{z\theta z}, \\
\varepsilon _{zr} &=&\frac{4\pi i}{\omega }W_{zzr},\varepsilon _{z\theta }=%
\frac{4\pi i}{\omega }W_{zz\theta },\varepsilon _{zz}=\frac{4\pi i}{\omega }%
W_{zzz}.
\end{eqnarray*}

\section{Dispersion relation in the case of the strong
collisional coupling of species}

We will further consider the case of the strong collisional coupling of
neutrals with ions and dust grains and dust grains with neutrals. We also
suppose that the collisional frequencies of electrons and ions with neutrals
are much larger than the perturbation frequency. Thus, the case under
consideration is the following:
\begin{equation}
\nu _{en}^{0}\gg \omega _{e},\nu _{i}^{0}\gg \omega _{i},\nu _{d}^{0}\gg
\omega _{d},\min \left\{ \nu _{ni}^{0},\nu _{nd}^{0}\right\} \gg \omega _{n},
\end{equation}

\noindent where
\[
\nu _{i}^{0}=\frac{\nu _{in}^{0}\nu _{nd}^{0}}{\nu _{n}^{0}},\nu _{d}^{0}=%
\frac{\nu _{dn}^{0}\nu _{ni}^{0}}{\nu _{n}^{0}}.
\]

\noindent The collisional frequencies $\nu _{j}$ [Eq. (36)] are
\[
\nu _{e}=\nu _{en}^{0},\nu _{i}=\nu _{i}^{0}-i\omega _{n}\frac{\nu
_{in}^{0}\nu _{ni}^{0}}{\nu _{n}^{02}},\nu _{d}=\nu _{d}^{0}-i\omega _{n}%
\frac{\nu _{dn}^{0}\nu _{nd}^{0}}{\nu _{n}^{02}}.
\]

\noindent Here we keep the corrections proportional to $\omega $ because the
main terms are cancelled when calculating some necessary expressions in the
case (44). For example, the value $\lambda _{\omega }$ has in this case the
form
\[
\lambda _{\omega }=\omega \left( \nu _{d}^{0}+\nu _{i}^{0}+\nu ^{0}\right)
=\omega a,
\]

\noindent where $\nu ^{0}=\nu _{in}^{0}\nu _{dn}^{0}/\nu _{n}^{0}$.

Let us find $W_{rr,\theta }$ and $W_{\theta r,\theta }$ under
conditions (44). We will consider, as in the steady state, that
the electrons are magnetized, $\omega _{ce}^{2}\gg \nu
_{en}^{02}.$ This condition is
generally satisfied. Using the condition of quasineutrality $%
\sum_{j}q_{j}n_{j0}=0$ and carrying out the calculations, we obtain the
following expressions for $W_{rr,\theta }$ and $W_{\theta r,\theta }$:
\begin{equation}
W_{rr}=W_{\theta \theta }=-iq_{e}n_{e0}\omega \sigma ,W_{r\theta
}=-W_{\theta r}=iq_{e}n_{e0}\omega \left( \sigma \frac{\nu _{en}^{0}}{\omega
_{ce}}-\varepsilon _{d}\right) ,
\end{equation}
where
\begin{eqnarray*}
\sigma &=&\nu _{d}^{0}a-\frac{\omega _{ci}}{\omega _{ce}}\nu _{d}^{02}g, \\
\varepsilon _{d} &=&\frac{q_{d}n_{d0}}{q_{e}n_{e0}}\omega _{ci}\nu
_{d}^{02}f.
\end{eqnarray*}

\noindent The values $g$ and $f$ are equal to
\[
g=\nu _{en}^{0}\frac{\nu _{nd}^{0}}{\nu _{ni}^{0}}\left( \frac{1}{\nu
_{dn}^{0}}+\frac{1}{\nu _{nd}^{0}}\right) ,f=\left( \frac{1}{\nu _{ni}^{0}}+%
\frac{1}{\nu _{d}^{0}}\right) .
\]

The values $s_{r,\theta ,zz}$ under conditions (44) and $\omega _{ce}^{2}\gg
\omega _{\nu e}^{2}$ are
\[
s_{rz}=i\nu _{d}^{0}\left[ \frac{\nu _{en}^{0}}{\omega _{ce}}\left(
v_{i0\theta }-v_{e0\theta }\right) -\left( v_{i0r}-v_{e0r}\right) \right]
-\omega \frac{q_{d}n_{d0}}{q_{e}n_{e0}}\nu _{d}^{0}fv_{i0r},
\]
\[
s_{\theta z}=i\nu _{d}^{0}\left[ -\frac{\nu _{en}^{0}}{\omega _{ce}}\left(
v_{i0r}-v_{e0r}\right) -\left( v_{i0\theta }-v_{e0\theta }\right) \right]
-\omega \frac{q_{d}n_{d0}}{q_{e}n_{e0}}\nu _{d}^{0}fv_{i0\theta },
\]
\begin{equation}
s_{zz}=-\omega \left[ \left( 1+g\right) \frac{\nu _{d}^{0}}{\nu _{en}^{0}}+%
\frac{q_{d}n_{d0}}{q_{e}n_{e0}}\nu _{d}^{0}f\right] .
\end{equation}

Using the expressions for $\varepsilon _{ij}$, $i,j=r,\theta ,z$, given in
\S\ 8 and taking into account Eqs. (45), (46), and the expressions for $%
W_{zrr,\theta ,z}$, $W_{z\theta r,\theta ,z}$, and $W_{zzr,\theta ,z}$ given
in the Appendix F, we find the following relations containing in the
dispersion relation (43):
\[
\varepsilon _{rr}\varepsilon _{zz}-\varepsilon _{rz}\varepsilon _{zr}=\tau
\beta \sigma \left( ib_{zz}-1\right) +i\beta ^{2}\frac{n_{z}^{2}}{c^{2}}%
\left( b_{zz}v_{e0r}-b_{rz}\right) \left( v_{i0r}-v_{e0r}\right) ,
\]
\[
-\varepsilon _{\theta r}\varepsilon _{zz}+\varepsilon _{\theta z}\varepsilon
_{zr}=-\tau \beta \left( \sigma \frac{\nu _{en}^{0}}{\omega _{ce}}%
-\varepsilon _{d}\right) \left( ib_{zz}-1\right) +i\beta ^{2}\frac{n_{z}^{2}%
}{c^{2}}\left( b_{\theta z}-b_{zz}v_{e0\theta }\right) \left(
v_{i0r}-v_{e0r}\right) ,
\]
\[
\varepsilon _{\theta \theta }\varepsilon _{zz}-\varepsilon _{\theta
z}\varepsilon _{z\theta }=\tau \beta \sigma \left( ib_{zz}-1\right) +i\beta
^{2}\frac{n_{z}^{2}}{c^{2}}\left( b_{zz}v_{e0\theta }-b_{\theta z}\right)
\left( v_{i0\theta }-v_{e0\theta }\right) ,
\]
\begin{equation}
-\varepsilon _{r\theta }\varepsilon _{zz}+\varepsilon _{rz}\varepsilon
_{z\theta }=\tau \beta \left( \sigma \frac{\nu _{en}^{0}}{\omega _{ce}}%
-\varepsilon _{d}\right) \left( ib_{zz}-1\right) +i\beta ^{2}\frac{n_{z}^{2}%
}{c^{2}}\left( b_{rz}-b_{zz}v_{e0r}\right) \left( v_{i0\theta }-v_{e0\theta
}\right) ,
\end{equation}
where
\[
\tau =\frac{\omega _{pe}^{2}}{\triangle }\frac{\omega _{ci}}{\omega _{ce}}%
,\beta =\frac{\omega _{pe}^{2}}{\omega \omega _{\nu e}}.
\]

\noindent Substituting the expressions (47) and $\varepsilon _{zz}$ in Eq.
(43), we obtain the following dispersion relation:
\[
\omega ^{6}\frac{\omega _{pe}^{4}}{\omega _{ce}^{2}\omega _{ci}^{2}}\left(
\sigma _{1}^{2}+\varepsilon _{d1}^{2}\right) +2\omega ^{4}\frac{\omega
_{pe}^{2}}{\omega _{ce}\omega _{ci}}\sigma _{1}k_{z}^{2}c^{2}+\omega
^{2}\left( k_{z}^{4}c^{4}-\frac{\omega _{pe}^{4}}{\omega _{ce}^{2}}%
k_{z}^{2}w_{0}^{2}\right)
\]
\begin{eqnarray}
&&-\frac{\omega _{pe}^{2}\omega _{ci}}{\omega _{ce}}\frac{1}{\left( \sigma
_{1}-\varepsilon _{d1}\frac{\nu _{en}^{0}}{\omega _{ce}}\right) }%
k_{z}^{4}c^{2}w_{0}^{2}+i\omega ^{3}\frac{\omega _{pe}^{4}}{\omega
_{ce}^{2}\omega _{ci}}\frac{\sigma _{1}\varepsilon _{d1}}{\left( \sigma
_{1}-\varepsilon _{d1}\frac{\nu _{en}^{0}}{\omega _{ce}}\right) }%
k_{z}^{2}w_{0}^{2} \\
&&+i\omega \frac{\omega _{pe}^{2}}{\omega _{ce}}\frac{\varepsilon _{d1}}{%
\left( \sigma _{1}-\varepsilon _{d1}\frac{\nu _{en}^{0}}{\omega _{ce}}%
\right) }k_{z}^{4}c^{2}w_{0}^{2}=0,  \nonumber
\end{eqnarray}

\noindent where
\begin{eqnarray*}
\sigma _{1} &=&\frac{a}{\nu _{d}^{0}}-\frac{\omega _{ci}}{\omega _{ce}}g, \\
\varepsilon _{d1} &=&\frac{q_{d}n_{d0}}{q_{e}n_{e0}}\omega _{ci}f, \\
w_{0}^{2} &=&\left( v_{i0r}-v_{e0r}\right) ^{2}+\left( v_{i0\theta
}-v_{e0\theta }\right) ^{2}.
\end{eqnarray*}

\noindent In Eq. (48) the relationship
\[
ib_{zz}-1=-\frac{\nu _{d}^{0}}{a}\left( \sigma _{1}-\varepsilon _{d1}\frac{%
\nu _{en}^{0}}{\omega _{ce}}\right)
\]

\noindent has been used. Below, we will find the unstable solutions of Eq.
(48), the growth rate of which is proportional to $w_{0}$.

\section{Solutions of the dispersion relation}

To solve the dispersion relation (48), we will consider the cases in which
the unstable perturbations have the long and short wavelengths.

\paragraph{\noindent \it {\textbf{10.1 Long wavelength instabilities}}\\}
\smallskip

\noindent \\
\noindent We neglect the terms proportional to $k_{z}^{4}$ in Eq. (48) under condition
\begin{equation}
\omega ^{2}\frac{\omega _{pe}^{2}}{\omega _{ce}\omega _{ci}}\sigma _{1}\gg
k_{z}^{2}c^{2}.
\end{equation}

\noindent Then the dispersion relation takes the form
\begin{equation}
\omega ^{4}\left( \sigma _{1}^{2}+\varepsilon _{d1}^{2}\right) -\omega
_{ci}^{2}k_{z}^{2}w_{0}^{2}+i\omega \omega _{ci}\frac{\sigma _{1}\varepsilon
_{d1}}{\left( \sigma _{1}-\varepsilon _{d1}\frac{\nu _{en}^{0}}{\omega _{ce}}%
\right) }k_{z}^{2}w_{0}^{2}=0.
\end{equation}

For frequencies in the range
\begin{equation}
\omega _{ci}\gg \omega \frac{\sigma _{1}\varepsilon _{d1}}{\left( \sigma
_{1}-\varepsilon _{d1}\frac{\nu _{en}^{0}}{\omega _{ce}}\right) },
\end{equation}

\noindent the unstable solution of Eq. (50) is the following:
\begin{equation}
\omega =i\frac{\left( \omega _{ci}k_{z}w_{0}\right) ^{1/2}}{\left( \sigma
_{1}^{2}+\varepsilon _{d1}^{2}\right) ^{1/4}}.
\end{equation}

In the case opposite to (51) the instability is caused by the presence of
dust grains:
\begin{equation}
\omega =\left[ \frac{\sigma _{1}\varepsilon _{d1}}{\left( \sigma
_{1}-\varepsilon _{d1}\frac{\nu _{en}^{0}}{\omega _{ce}}\right) }\frac{%
\omega _{ci}k_{z}^{2}w_{0}^{2}}{\left( \sigma _{1}^{2}+\varepsilon
_{d1}^{2}\right) }\right] ^{1/3}\exp \left( -i\frac{\pi }{6}+i\frac{2p\pi }{3%
}\right) ,
\end{equation}

\noindent where $p=0,1,2.$ Note, that the sign of $\varepsilon _{d1}$
depends on the sign of $q_{d}$.

\paragraph{\noindent \it {\textbf {10.2 Short wavelength instabilities}}\\}
\smallskip

\noindent \\
\noindent Now consider the case in which
\begin{equation}
\omega ^{2}\frac{\omega _{pe}^{2}}{\omega _{ce}\omega _{ci}}\left( \sigma
_{1}^{2}+\varepsilon _{d1}^{2}\right) ^{1/2}\ll k_{z}^{2}c^{2}.
\end{equation}

\noindent Then the equation (48) will be the following:
\begin{equation}
\omega ^{2}-\frac{\omega _{pe}^{2}\omega _{ci}}{\omega _{ce}}\frac{1}{\left(
\sigma _{1}-\varepsilon _{d1}\frac{\nu _{en}^{0}}{\omega _{ce}}\right) }%
\frac{w_{0}^{2}}{c^{2}}+i\omega \frac{\omega _{pe}^{2}}{\omega _{ce}}\frac{%
\varepsilon _{d1}}{\left( \sigma _{1}-\varepsilon _{d1}\frac{\nu _{en}^{0}}{%
\omega _{ce}}\right) }\frac{w_{0}^{2}}{c^{2}}=0.
\end{equation}

For frequencies in the range
\begin{equation}
\omega _{ci}\gg \omega \varepsilon _{d1}
\end{equation}

\noindent we find the following solution of Eq. (55):
\begin{equation}
\omega ^{2}=\frac{\omega _{pe}^{2}\omega _{ci}}{\omega _{ce}}\frac{1}{\left(
\sigma _{1}-\varepsilon _{d1}\frac{\nu _{en}^{0}}{\omega _{ce}}\right) }%
\frac{w_{0}^{2}}{c^{2}}.
\end{equation}

\noindent We see that when $\sigma _{1}>\varepsilon _{d1}(\nu
_{en}^{0}/\omega _{ce})$ or $\varepsilon _{d1}>0$ ($q_{d}<0$) there is the
streaming instability.

In the case opposite to (56),
\begin{equation}
\omega _{ci}\ll \omega \varepsilon _{d1},
\end{equation}

\noindent we obtain
\begin{equation}
\omega =-i\frac{\omega _{pe}^{2}}{\omega _{ce}}\frac{\varepsilon _{d1}}{%
\left( \sigma _{1}-\varepsilon _{d1}\frac{\nu _{en}^{0}}{\omega _{ce}}%
\right) }\frac{w_{0}^{2}}{c^{2}}.
\end{equation}

\noindent This solution describes the instability when $\varepsilon _{d1}>0$
or $\varepsilon _{d1}(\nu _{en}^{0}/\omega _{ce})>\sigma _{1}.$

\section{Discussion}

In the equilibrium and perturbation states we have not taken into account
the collisions between ions and dust grains, supposing that $\nu _{in}\gg
\nu _{id}$ and $\nu _{dn}\gg \nu _{di}$. These inequalities can be written
in the form $<\sigma v>_{in}\gg $ $\left( m_{i}\rho _{d}/m_{d}\rho
_{n}\right) <\sigma v>_{id}$and $<\sigma v>_{dn}\gg (\rho _{i}/\rho
_{n})<\sigma v>_{id}$, where $\rho _{i,d,n}=m_{i,d,n}n_{i,d,n}$ $(m_{i}\gg
m_{n})$. Such conditions can be satisfied, if the density of neutrals is
sufficiently high.

Now we estimate the stationary velocities of species in the case $\nu
_{in}\gg \nu _{ni}$ and $\nu _{dn}\gg \nu _{ni},\nu _{nd}$. These conditions
can be satisfied in the weakly ionized protostellar and protoplanetary
disks. Then, neglecting the contribution of the radial electric field under
condition $E_{0r}/B_{0}\ll (\rho _{n}/\rho _{i})(\Omega _{K}/\omega
_{ci})(v_{K}/c)$, we obtain from Eq. (25)
\begin{equation}
v_{n0\theta }^{0}\simeq \frac{\nu _{in}}{\omega _{ci}}\frac{\Omega _{K}}{\nu
_{ni}}v_{K},
\end{equation}

\noindent where $v_{K}$ and $\Omega _{K}$ are the Keplerian velocity and
frequency, correspondingly. Using Eq. (60), we can write the condition of
the strong collisional coupling of neutrals with ions (20) in the form
\begin{equation}
\frac{\omega _{ci}^{2}}{\omega _{ci}^{2}+\nu _{in}^{2}}\nu _{ni}^{2}\gg
\Omega _{K}^{2}.
\end{equation}

\noindent We see from Eqs. (60) and (61) that $v_{n0\theta }^{0}\ll v_{K}$.
As long as $v_{d0\theta }^{0}\simeq v_{n0\theta }^{0}$ [see Eq. (27)] and $%
v_{i0\theta }^{0}\simeq v_{n0\theta }^{0}$, the species rotate in the
magnetized regions of the disk with the velocity much smaller than the
Keplerian velocity. This result agrees with observations (Donati et al.
2005). From Eqs. (26) and (28) we obtain the radial velocities of neutrals
and dust grains under conditions at hand
\begin{equation}
v_{n0r}^{0}\simeq v_{d0r}^{0}\simeq -\frac{\Omega _{K}}{\nu _{ni}}v_{K}.
\end{equation}

\noindent Comparing Eqs. (60) and (62), we have $v_{n,d0\theta }^{0}\simeq
-(\nu _{in}/\omega _{ci})v_{n,d0r}^{0}$. Thus, for the magnetized
(unmagnetized) ions, $\omega _{ci}\gg (\ll )\nu _{in}$, the radial velocity
of neutrals and dust grains is larger (smaller) than their azimuthal
velocity. We see from Eq. (62) that the radial velocity of neutrals and dust
grains is directed inward of the disk.

It is followed from Eq. (29) that the azimuthal electron velocity $%
v_{e0\theta }\ll v_{i0\theta }^{0}$ and the radial electron velocity $%
v_{e0r}\simeq (\nu _{en}/\omega _{ce})v_{n0\theta }^{0}$.

The conditions (21) and (23) of the strong collisional coupling of dust
grains with neutrals can be written in the form
\begin{equation}
\nu _{dn}\gg \Omega _{K}\left[ 1+\frac{\Omega _{K}}{\nu _{ni}}\left( 1+\frac{%
\nu _{in}^{2}}{\omega _{ci}^{2}}\right) \right] .
\end{equation}

Let us consider some specific parameters for the protoplanetary disk with
the solar mass central star at $r=1$\ AU, where $B_{0}\sim 0.1$\ G and $%
\Omega _{K}\sim 2\times 10^{-7}$\ s$^{-1}$\ (e.g., Desch 2004). We take $%
m_{i}=30m_{p}$ and $m_{n}=2.33m_{p}$, $m_{p}$\ is the proton mass. Then the
ion and electron cyclotron frequencies will be equal to $\omega _{ci}\sim 32$%
\ s$^{-1}$ ($q_{i}=-q_{e}$) and $|\omega _{ce}|\sim 1.76\times 10^{6}$ s$%
^{-1}$ (the sign $||$ denotes an absolute value). The rate coefficients for
momentum transfer by elastic scattering of ions and electrons with neutrals
are $<\sigma \nu >_{in}=1.9\times 10^{-9}$ cm$^{3}$ s$^{-1}$ and $<\sigma
\nu >_{en}=4.5\times 10^{-9}(T/30$ K$)^{1/2}$ cm$^{3}$ s$^{-1}$ (Draine et
al. 1983). We take $T=300$ K. Then the condition for neglecting collisions
of neutrals with electrons has the form $\omega _{ci}^{2}/(\omega
_{ci}^{2}+\nu _{in}^{2})\gg 1.89\times 10^{-3}(n_{e0}/n_{i0})$ (see the end
of \S\ 3). We will consider the case in which $\nu _{in}^{2}\gg \omega
_{ci}^{2}$ and $n_{e0}/n_{i0}\sim 10^{-2}$ (Desch 2004). Then the value $\nu
_{in}$ must satisfy the condition $\nu _{in}<7.36\times 10^{3}$ s$^{-1}$.
This inequality is satisfied for the neutral mass density $\rho
_{n}<2.1\times 10^{-10}$ g cm$^{-3}$ or $n_{n0}<5.38\times 10^{13}$ cm$^{-3}$%
. We will take the ionization degree as $n_{i0}=10^{-9}n_{n0}$ (Desch 2004).
If we set $\rho _{n}=10^{-10}$ g cm$^{-3}$, then we obtain $\nu
_{in}=3.5\times 10^{3}$ s$^{-1}$, $\nu _{ni}=4.5\times 10^{-5}$ s$^{-1}$, $%
\nu _{en}=3.65\times 10^{5}$ s$^{-1}$, $\nu _{ne}=8.53\times 10^{-10}$ s$%
^{-1}$.

The azimuthal velocity of species using the parameters given above is equal
to $v_{i,d,n0\theta }^{0}\simeq 0.49v_{K},$ where $v_{K}\simeq 30$ km s$%
^{-1} $ [see Eq. (60)]. The orbiting frequency of dust grains $\Omega _{d}=$
$0.49\Omega _{K}=0.98\times 10^{-8}$ s$^{-1}$. The dust grains will be
unmagnetized, $\omega _{cd}\ll \Omega _{d}$, if their mass $m_{d}\gg 8\times
10^{-15}$ g $\simeq 4.8\times 10^{9}m_{p}$ (for $q_{d}=\pm q_{e}$). The dust
grains, for example, with density of the material grains are made $\sigma
_{d}\sim 3$ g cm$^{-3}$ and with radius $r_{d}>8.6\times 10^{-2}$ $\mu $m
satisfy the condition of unmagnetization. At $r_{d}=0.1$ $\mu $m, the mass
of the grain is equal to $m_{d}\simeq 1.26\times 10^{-14}$ g. The
collisional frequency of dust grains with neutrals has the form $\nu
_{dn}\simeq 6.7\rho _{n}r_{d}^{2}v_{Tn}/m_{d}$ (e.g., Wardle and Ng 1999).
Using parameters given above, we obtain $\nu _{dn}\simeq 0.53$ s$^{-1}$. For
the solar abundance value $\rho _{d}\simeq 10^{-2}$ $\rho _{n}$ we have $\nu
_{nd}\simeq 0.53\times 10^{-2}$ s$^{-1}$.

The radial velocity of neutrals and dust grains is equal to $%
|v_{d,n0r}|\simeq 133$ m s$^{-1}$ [see Eq. (62)]. This velocity is directed
inward. The conditions (61) and (63) are satisfied.

Above, we have considered only one set of disk parameters. Any other
parameters can be treated also.

In the case (44) of the strong collisional coupling between ions, neutrals,
and dust grains the perturbed velocity of dust grains is of the order of the
perturbed velocity of ions, $\mathbf{v}_{d1}\sim \mathbf{v}_{i1}$ [see Eqs.
(39) and (40)]. Thus, the dust grains can participate in the fast
electromagnetic perturbations generated by the electrons and ions via
collisions with neutrals and acquire the large velocities.

Let us now consider the obtained unstable solutions. For conditions used in
this section the growth rate $\gamma =$Im $\omega $ of the solution (52) can
be estimated in the case $\sigma _{1}\geq \varepsilon _{d1}$ as
\begin{equation}
\gamma \sim \left( k_{z}r\right) ^{1/2}\Omega _{K}.
\end{equation}

\noindent This growth rate is much larger than the Keplerian frequency as
long as $k_{z}r\gg 1$. The wave number $k_{z}$ must satisfy the condition
[see the inequality (49)]
\[
k_{z}r\ll \frac{|q_{e}|n_{e0}}{q_{i}n_{i0}}\frac{\rho _{n}}{\rho _{i}}\frac{%
v_{K}^{2}}{c_{Ai}^{2}},
\]

\noindent where $c_{Ai}=(\omega _{ci}/\omega _{pi})c$ is the ion Alfv\'{e}n
velocity. For the parameters given above we have $n_{i0}=2.56\times 10^{4}$
cm$^{-3}$. In this case $\omega _{pi}\simeq 3.84\times 10^{4}$ s$^{-1}$ and,
accordingly, $c_{Ai}\simeq 250$ km s$^{-1}$. So far as $\gamma \ll \nu _{ni}$%
, we have the second condition: $\left( k_{z}r\right) ^{1/2}\ll \nu
_{ni}/\Omega _{K}.$ \noindent It is interesting to note that the same growth
rate also exists in the collisionless regime (Nekrasov 2007). The growth
rate (53) is larger than the growth rate defined by Eq. (64). This
instability is possible when the density of dust grains is sufficiently
large:
\[
\frac{\nu _{in}}{\omega _{ci}}\geq \frac{q_{d}n_{d0}}{|q_{e}|n_{e0}}\gg
\frac{\nu _{ni}}{\Omega _{K}}\left( k_{z}r\right) ^{-1/2}.
\]

The short wavelength perturbations considered in section (10.2) have the
following growth rates for conditions given above. An estimation for the
growth rate (57) is
\begin{equation}
\gamma \sim \left( \frac{|q_{e}|n_{e0}}{q_{i}n_{i0}}\frac{\rho _{n}}{\rho
_{i}}\right) ^{1/2}\frac{v_{K}}{c_{Ai}}\Omega _{K}.
\end{equation}

\noindent This growth rate is considerably larger than the Keplerian
frequency in weakly ionized disks. For the parameters given above we have $%
\gamma \sim $ $2.1\times 10^{-5}$ s$^{-1}$. The condition (54) at $\sigma
_{1}\geq \varepsilon _{d1}$ is the following:
\[
k_{z}r\gg \frac{|q_{e}|n_{e0}}{q_{i}n_{i0}}\frac{\rho _{n}}{\rho _{i}}\frac{%
v_{K}^{2}}{c_{Ai}^{2}}.
\]
The growth rate (59) is larger than (65) and has the form $(q_{d}<0)$
\[
\gamma \sim \frac{|q_{d}|n_{d0}}{q_{i}n_{i0}}\frac{\rho _{n}}{\rho _{i}}%
\frac{\Omega _{K}^{2}}{\nu _{ni}}\frac{v_{K}^{2}}{c_{Ai}^{2}}.
\]

\noindent For this instability the following condition must be satisfied
[see the inequality (58)]: $q_{d}n_{d0}/q_{e}n_{e0}\gg \nu _{ni}/\gamma \gg
1.$

Thus, in the collisional accretion disks there are possible the streaming
instabilities involving the dynamics of neutrals and dust grains with growth
rates much larger that the Keplerian frequency.

In the present paper, we did not take into account the ionization and
recombination processes in the continuity and momentum equations which, in
general, can be included (see, e.g., Pinto et al. 2008; Li et al. 2008). The
main sources of ionization of accretion disks are Galactic cosmic rays and
X-rays from the corona of the central star. These and other chemical
processes can play an important physical role in the evolution of the disk.
For example, the dead zones of the protostellar and protoplanetary disks may
be enlivened due to turbulent transport of metallic ions, which are charged
due to interaction with ionized gas in the surface layers, from the surface
layers into these regions (Ilgner and Nelson 2008).

The ionization/recombination processes in the stationary continuity
equations for species determine the ionization degree of medium. For the
weakly ionized gases with large collisional frequencies, as it is the case
for us, the contribution from the ionization source terms to the momentum
equations can be ignored (e.g., Li et al. 2008). The chemical processes do
not influence the dynamics of medium, for example, in the case of the ideal
magnetohydrodynamics. In our case of multicomponent disks with strong
collisional coupling of species, when the dynamics of each species is
considered separately, the contribution of the ionization source term to the
perturbed continuity equation can be neglected under condition $\omega >\xi
(n_{n0}/n_{i0})$, where $\xi $ is the ionization rate per H atom. The
typical growth rates of the streaming instabilities considered in the
present paper is $\gamma \sim 10^{-5}$ s$^{-1}$. If we take $%
n_{n0}/n_{i0}\sim 10^{9}$, then we can neglect the ionization for $\xi
<10^{-14}$ s$^{-1}$. Note, for example, that in the paper (Ilgner and Nelson
2008) the range of $\xi \sim 10^{-15}-10^{-19}$ s$^{-1}$ is considered. For
cosmic rays one obtains $\xi \sim 3\times 10^{-17}$ s$^{-1}$ (e.g., Li et
al. 2008). Thus, for not too high ionization rates and fast instabilities
studied in the present paper, the ionization/recombination processes can be
ignored.

\section{Conclusion}

In the present paper, the electromagnetic streaming instabilities of
multicomponent collisional accretion disks have been studied. We have
explored regions of the disk where the medium is sufficiently ionized and
there is strong collisional coupling of neutrals with ions and dust grains
simultaneously. We have included the effect of perturbation of collisional
frequencies due to density perturbations of charged species and neutrals.
This effect emerges when there are different background velocities of
species, as occurs in accretion disks.

We have investigated in detail the steady state and found the
azimuthal and radial velocities of species.

The general solutions for the perturbed velocities of species with
collisional and thermal effects have been obtained. We have shown that dust
grains can be involved in the fast electromagnetic perturbations induced by
ions and electrons, through strong collisions with neutrals which have also
strong collisional coupling with ions. The dust grains have been found to
acquire the perturbed velocity of ions. This effect is important for their
collisional coagulation and sticking.

We have derived the dispersion relation for the vertical perturbations and
found the unstable solutions due to different background velocities of
electrons and ions. It has been shown that the growth rates of these
streaming instabilities can be much larger than the Keplerian frequency.

Electromagnetic streaming instabilities, with induced dynamics of neutrals
can be a source of turbulence in sufficiently ionized regions of collisional
accretion disks.

\paragraph{Acknowledgments\\}

I thank the anonymous referee for his/her constructive comments.

\paragraph{APPENDIX A\\}

\smallskip
\paragraph{Solutions for $\mathbf{v}_{n1k}$ and $n_{n1k}$\\}
\renewcommand{\theequation}{A-\arabic{equation}}
\setcounter{equation}{0} The Fourier components $\mathbf{v}_{n1k}$
and $n_{n1k}$ are the following (for simplicity, the index $k$ is
omitted here and below):
\begin{eqnarray*}
D_{n}v_{n1r} &=&iR_{n1r}\omega _{\nu n}\left( 1-\delta _{\theta \theta
n}-\delta _{\mid \mid n}\right) +iR_{n1\theta }\left[ \omega _{\nu n}\delta
_{r\theta n}+i2\Omega _{n}\left( 1-\delta _{\mid \mid n}\right) \right] \\
&&+iR_{n1z}\left( \omega _{\nu n}\delta _{rzn}+i2\Omega _{n}\delta _{\theta
zn}\right) ,
\end{eqnarray*}
\begin{eqnarray*}
D_{n}v_{n1\theta } &=&iR_{n1r}\left[ \omega _{\nu n}\delta _{r\theta n}-i%
\frac{\kappa _{n}^{2}}{2\Omega _{n}}\left( 1-\delta _{\mid \mid n}\right)
\right] +iR_{n1\theta }\omega _{\nu n}\left( 1-\delta _{rrn}-\delta _{\mid
\mid n}\right) \\
&&+iR_{n1z}\left( \omega _{\nu n}\delta _{\theta zn}-i\frac{\kappa _{n}^{2}}{%
2\Omega _{n}}\delta _{rzn}\right) ,
\end{eqnarray*}
\begin{eqnarray*}
D_{n}v_{n1z} &=&iR_{n1r}\left( \omega _{\nu n}\delta _{rzn}-i\frac{\kappa
_{n}^{2}}{2\Omega _{n}}\delta _{\theta zn}\right) +iR_{n1\theta }\left(
\omega _{\nu n}\delta _{\theta zn}+i2\Omega _{n}\delta _{rzn}\right) \\
&&+iR_{n1z}\left[ \omega _{\nu n}\left( 1-\delta _{\perp n}\right) -\frac{%
\kappa _{n}^{2}}{\omega _{\nu n}}-i\left( 2\Omega _{n}-\frac{\kappa _{n}^{2}%
}{2\Omega _{n}}\right) \delta _{r\theta n}\right] ,
\end{eqnarray*}
\[
\frac{\omega _{n}}{\omega _{\nu n}}D_{n}\frac{n_{n1}}{n_{n0}}=iR_{n1r}\left(
k_{r}-i\frac{\kappa _{n}^{2}}{2\Omega _{n}\omega _{\nu n}}k_{\theta }\right)
+iR_{n1\theta }\left( k_{\theta }+i\frac{2\Omega _{n}}{\omega _{\nu n}}%
k_{r}\right) +iR_{n1z}\left( k_{z}-\frac{\kappa _{n}^{2}}{\omega _{\nu n}^{2}%
}k_{z}\right) .
\]

\noindent The following notations are introduced here:
\[
D_{n}=\omega _{\nu n}^{2}\left( 1-\delta _{n}\right) -\kappa _{n}^{2}\left(
1-\delta _{\mid \mid n}\right) -i\omega _{\nu n}\left( 2\Omega _{n}-\frac{%
\kappa _{n}^{2}}{2\Omega _{n}}\right) \delta _{r\theta n},
\]
\[
\omega _{\nu n}=\omega _{n}+i\nu _{n}^{0},\omega _{n}=\omega -\mathbf{kv}%
_{n0},
\]
\[
\Omega _{n}=\frac{v_{n0\theta }}{r},\kappa _{n}^{2}=(2\Omega _{n}/r)\partial
(r^{2}\Omega _{n})/\partial r,k_{\theta }=\frac{m}{r},
\]
\[
\delta _{lmn}=\frac{k_{l}k_{m}v_{Tn}^{2}}{\omega _{n}\omega _{\nu n}},\delta
_{\perp n}=\delta _{rrn}+\delta _{\theta \theta n},\delta _{\mid \mid
n}=\delta _{zzn},\delta _{n}=\delta _{\perp n}+\delta _{\mid \mid n},
\]

\noindent where $l,m=r,\theta ,z$.

\paragraph{APPENDIX B}

\smallskip
\paragraph{Solutions for $\mathbf{v}_{j1k}$ and $n_{j1k}$\\}
\renewcommand{\theequation}{B-\arabic{equation}}
\setcounter{equation}{0} The Fourier components $\mathbf{v}_{j1k}$
and $n_{j1k}$ are:
\begin{eqnarray*}
D_{j}v_{j1r} &=&iF_{j1r}\omega _{\nu j}\left( 1-\delta _{\theta \theta
j}-\delta _{\mid \mid j}\right) +iF_{j1\theta }\left[ \omega _{\nu j}\delta
_{r\theta j}+i\omega _{j1}\left( 1-\delta _{\mid \mid j}\right) \right] \\
&&+iF_{j1z}\left( \omega _{\nu j}\delta _{rz\perp j}+i\omega _{j1}\delta
_{\theta z\perp j}\right) ,
\end{eqnarray*}
\begin{eqnarray*}
D_{j}v_{j1\theta } &=&iF_{j1r}\left[ \omega _{\nu j}\delta _{r\theta
j}-i\omega _{j2}\left( 1-\delta _{\mid \mid j}\right) \right] +iF_{j1\theta
}\omega _{\nu j}\left( 1-\delta _{rrj}-\delta _{\mid \mid j}\right) \\
&&+iF_{j1z}\left( \omega _{\nu j}\delta _{\theta z\perp j}-i\omega
_{j2}\delta _{rz\perp j}\right) ,
\end{eqnarray*}
\begin{eqnarray*}
D_{j}v_{j1z} &=&iF_{j1r}\left( \omega _{\nu j}\delta _{rz\mid \mid
j}-i\omega _{j2}\delta _{\theta z\mid \mid j}\right) +iF_{j1\theta }\left(
\omega _{\nu j}\delta _{\theta z\mid \mid j}+i\omega _{j1}\delta _{rz\mid
\mid j}\right) \\
&&+iF_{j1z}\left[ \omega _{\nu j}\left( 1-\delta _{\perp j}\right) -\frac{%
\omega _{j1}\omega _{j2}}{\omega _{\nu j}}-i\left( \omega _{j1}-\omega
_{j2}\right) \delta _{r\theta j}\right] ,
\end{eqnarray*}
\[
\frac{\omega _{j}}{\omega _{\nu j}}D_{j}\frac{n_{j1}}{n_{j0}}=iF_{j1r}\left(
k_{r}-i\frac{\omega _{j2}}{\omega _{\nu j}}k_{\theta }\right) +iF_{j1\theta
}\left( k_{\theta }+i\frac{\omega _{j1}}{\omega _{\nu j}}k_{r}\right)
+iF_{j1z}\left( k_{z}-\frac{\omega _{j1}\omega _{j2}}{\omega _{\nu j}^{2}}%
k_{z}\right) .
\]

\noindent The following notations are introduced here:
\[
D_{j}=\omega _{\nu j}^{2}\left( 1-\delta _{\perp j}-\delta _{\mid \mid
j}\right) -\omega _{j1}\omega _{j2}\left( 1-\delta _{\mid \mid j}\right)
-i\omega _{\nu j}\left( \omega _{j1}-\omega _{j2}\right) \delta _{r\theta
j},
\]
\[
F_{j1r}=G_{j1r}+Q_{j1r}=\frac{q_{j}}{m_{j}}E_{1r}\left( 1-n_{\theta }\frac{%
v_{j0\theta }}{c}\right) +\frac{q_{j}}{m_{j}}E_{1\theta }n_{r}\frac{%
v_{j0\theta }}{c}+Q_{j1r},
\]
\[
F_{j1\theta }=G_{j1\theta }+Q_{j1\theta }=\frac{q_{j}}{m_{j}}E_{1\theta
}\left( 1-n_{r}\frac{v_{j0r}}{c}\right) +\frac{q_{j}}{m_{j}}E_{1r}n_{\theta }%
\frac{v_{j0r}}{c}+Q_{j1\theta },
\]
\[
F_{j1z}=G_{j1z}+Q_{j1z}=\frac{q_{j}}{m_{j}}E_{1z}\frac{\omega _{j}}{\omega }+%
\frac{q_{j}}{m_{j}}\mathbf{E}_{1}\mathbf{v}_{j0}\frac{n_{z}}{c}+Q_{j1z}.
\]
\[
\omega _{\nu j}=\omega _{j}+i\nu _{j},\omega _{j}=\omega -\mathbf{kv}_{j0},
\]
\[
\omega _{j1}=\omega _{cj}+2\Omega _{j},\omega _{j2}=\omega _{cj}+\kappa
_{j}^{2}/2\Omega _{j},
\]
\[
\Omega _{j}=\frac{v_{j0\theta }}{r},\kappa _{j}^{2}=(2\Omega _{j}/r)\partial
(r^{2}\Omega _{j})/\partial r,
\]
\[
\delta _{lmj}=\frac{k_{l}k_{m}v_{\perp j}^{2}}{\omega _{j}\omega _{\nu j}}%
,\delta _{lz\perp j}=\frac{k_{l}k_{z}v_{\perp j}^{2}}{\omega _{j}\omega
_{\nu j}},\delta _{lz\mid \mid j}=\frac{k_{l}k_{z}v_{\mid \mid j}^{2}}{%
\omega _{j}\omega _{\nu j}},\delta _{\perp j}=\delta _{rrj}+\delta _{\theta
\theta j},\delta _{\mid \mid j}=\frac{k_{z}^{2}v_{\mid \mid j}^{2}}{\omega
_{j}\omega _{\nu j}},
\]

\noindent where $l,m=r,\theta $ and
\[
v_{\perp j}^{2}=\frac{\gamma _{\perp j}T_{\perp 0j}}{m_{j}},v_{\mid \mid
j}^{2}=\frac{\gamma _{\mid \mid j}T_{\mid \mid 0j}}{m_{j}}.
\]

\noindent Here $T_{\perp 0j}$ and $T_{\mid \mid 0j}$ are the unperturbed
temperatures of species $j$ across ($\perp $) and along ($\mid \mid $) the
magnetic field $\mathbf{B}_{0}$.

\paragraph{\noindent APPENDIX C}

\smallskip
\paragraph{The system of equations for perturbations of ions and dust grains
\\}
\renewcommand{\theequation}{C\arabic{equation}}
\setcounter{equation}{0}
 Solutions given in the Appendix B at taking into account
collisions of neutrals with ions and dust grains simultaneously
and neglecting the thermal effects, result in the following system
of equations for perturbations of ions and dust grains:
\[
D_{i}v_{i1r}+a_{dr}v_{d1r}+a_{d\theta 1}v_{d1\theta }-b_{dr}\frac{n_{d1}}{%
n_{d0}}=-H_{i1r},
\]
\[
D_{i}v_{i1\theta }-a_{d\theta 2}v_{d1r}+a_{dr}v_{d1\theta }-b_{d\theta }%
\frac{n_{d1}}{n_{d0}}=-H_{i1\theta },
\]
\[
\omega _{\nu i}^{2}v_{i1z}+a_{dr}v_{d1z}=iG_{i1z}\omega _{\nu i},
\]
\begin{equation}
\left( k_{r}a_{dr}-k_{\theta }a_{d\theta 2}\right) v_{d1r}+\left( k_{\theta
}a_{dr}+k_{r}a_{d\theta 1}\right) v_{d1\theta }+\omega _{i}D_{i}\frac{n_{i1}%
}{n_{i0}}-\mathbf{kb}_{d}\frac{n_{d1}}{n_{d0}}=-\mathbf{kH}_{i1},
\end{equation}
\[
D_{d}v_{d1r}+a_{ir}v_{i1r}+a_{i\theta 1}v_{i1\theta }-b_{ir}\frac{n_{i1}}{%
n_{i0}}=-H_{d1r},
\]
\[
D_{d}v_{d1\theta }-a_{i\theta 2}v_{i1r}+a_{ir}v_{i1\theta }-b_{i\theta }%
\frac{n_{i1}}{n_{i0}}=-H_{d1\theta },
\]
\[
\omega _{\nu d}^{2}v_{d1z}+a_{ir}v_{i1z}=iG_{d1z}\omega _{\nu d},
\]
\begin{equation}
\left( k_{r}a_{ir}-k_{\theta }a_{i\theta 2}\right) v_{i1r}+\left( k_{\theta
}a_{ir}+k_{r}a_{i\theta 1}\right) v_{i1\theta }+\omega _{d}D_{d}\frac{n_{d1}%
}{n_{d0}}-\mathbf{kb}_{i}\frac{n_{i1}}{n_{i0}}=-\mathbf{kH}_{d1}.
\end{equation}

\noindent Here $D_{i,d}=\omega _{\nu i,d}^{2}-\omega _{i,d1}\omega _{i,d2}.$
The following notations are introduced above:
\begin{eqnarray*}
H_{i1r} &=&-i\omega _{\nu i}G_{i1r}+\omega _{i1}G_{i1\theta },H_{i1\theta
}=-\omega _{i2}G_{i1r}-i\omega _{\nu i}G_{i1\theta }, \\
H_{d1r} &=&-i\omega _{\nu d}G_{d1r}+\omega _{d1}G_{d1\theta },H_{d1\theta
}=-\omega _{d2}G_{d1r}-i\omega _{\nu d}G_{d1\theta }. \\
H_{i1z} &=&-i\frac{D_{i}}{\omega _{\nu i}}G_{i1z}+a_{dr}\frac{D_{i}}{\omega
_{\nu i}^{2}}v_{d1z},H_{d1z}=-i\frac{D_{d}}{\omega _{\nu d}}G_{d1z}+a_{ir}%
\frac{D_{d}}{\omega _{\nu d}^{2}}v_{i1z}.
\end{eqnarray*}
\begin{eqnarray*}
a_{ir} &=&\omega _{\nu d}\mu _{i},a_{i\theta 1}=i\omega _{d1}\mu
_{i},a_{i\theta 2}=i\omega _{d2}\mu _{i},b_{ir}=-a_{ir}w_{r}-a_{i\theta
1}w_{\theta },b_{i\theta }=-a_{ir}w_{\theta }+a_{i\theta 2}w_{r}, \\
a_{dr} &=&\omega _{\nu i}\mu _{d},a_{d\theta 1}=i\omega _{i1}\mu
_{d},a_{d\theta 2}=i\omega _{i2}\mu _{d},b_{dr}=a_{dr}w_{r}+a_{d\theta
1}w_{\theta },b_{d\theta }=a_{dr}w_{\theta }-a_{d\theta 2}w_{r},
\end{eqnarray*}

\noindent where $\mathbf{w}=\mathbf{v}_{i0}-v_{d0}$ and
\[
\mu _{i}=\frac{\nu _{dn}^{0}\nu _{ni}^{0}}{\omega _{\nu n}},\mu _{d}=\frac{%
\nu _{in}^{0}\nu _{nd}^{0}}{\omega _{\nu n}}.
\]

\paragraph{APPENDIX D}

\smallskip
\paragraph{Solutions of the system of equations given in Appendix C in the
case $k_{z}\neq 0,$ $k_{r}=k_{\theta }=0$\\}
\renewcommand{\theequation}{D\arabic{equation}}
\setcounter{equation}{0}
The exact solutions of the system of equations given in the Appendix C for $%
v_{i1r,\theta }$ in the case $k_{z}\neq 0,$ $k_{r}=k_{\theta }=0$ have the
form
\begin{eqnarray*}
Av_{i1r} &=&-\left( \omega _{\nu d}\lambda _{\omega }+i\omega _{\nu i}\omega
_{d1}\omega _{d2}\right) G_{i1r}-\left( D_{d}\omega _{i1}+\mu _{i}\mu
_{d}\omega _{d1}\right) G_{i1\theta } \\
&&+\mu _{d}\left( \lambda _{\omega }-i\omega _{i1}\omega _{d2}\right)
G_{d1r}+\mu _{d}\left( \omega _{\nu d}\omega _{i1}+\omega _{\nu i}\omega
_{d1}\right) G_{d1\theta } \\
&&-\mu _{i}\mu _{d}\omega ^{-1}D_{i}^{-1}\left[ w_{r}\left( \omega _{\nu
i}\omega _{\nu d}-\mu _{i}\mu _{d}+\omega _{i1}\omega _{d2}\right)
+iw_{\theta }\left( \omega _{\nu d}\omega _{i1}+\omega _{\nu i}\omega
_{d1}\right) \right] k_{z}H_{i1z}
\end{eqnarray*}
\begin{equation}
-\mu _{d}\omega ^{-1}D_{d}^{-1}\left[ w_{r}\left( D_{d}\omega _{\nu i}-\mu
_{i}\mu _{d}\omega _{\nu d}\right) +iw_{\theta }\left( D_{d}\omega _{i1}+\mu
_{i}\mu _{d}\omega _{d1}\right) \right] k_{z}H_{d1z},
\end{equation}
\begin{eqnarray*}
Av_{i1\theta } &=&\left( D_{d}\omega _{i2}+\mu _{i}\mu _{d}\omega
_{d2}\right) G_{i1r}-\left( \omega _{\nu d}\lambda _{\omega }+i\omega _{\nu
i}\omega _{d1}\omega _{d2}\right) G_{i1\theta } \\
&&-\mu _{d}\left( \omega _{\nu d}\omega _{i2}+\omega _{\nu i}\omega
_{d2}\right) G_{d1r}+\mu _{d}\left( \lambda _{\omega }-i\omega _{i2}\omega
_{d1}\right) G_{d1\theta } \\
&&-\mu _{i}\mu _{d}\omega ^{-1}D_{i}^{-1}\left[ -iw_{r}\left( \omega _{\nu
d}\omega _{i2}+\omega _{\nu i}\omega _{d2}\right) +w_{\theta }\left( \omega
_{\nu i}\omega _{\nu d}-\mu _{i}\mu _{d}+\omega _{i2}\omega _{d1}\right)
\right] k_{z}H_{i1z}
\end{eqnarray*}
\begin{equation}
-\mu _{d}\omega ^{-1}D_{d}^{-1}\left[ -iw_{r}\left( D_{d}\omega _{i2}+\mu
_{i}\mu _{d}\omega _{d2}\right) +w_{\theta }\left( D_{d}\omega _{\nu i}-\mu
_{i}\mu _{d}\omega _{\nu d}\right) \right] k_{z}H_{d1z},
\end{equation}

\noindent where the value $\lambda _{\omega }$ is equal to
\[
i\lambda _{\omega }=\omega _{\nu i}\omega _{\nu d}-\mu _{i}\mu _{d}.
\]

The value $A$ (the determinant of the system is equal to $A\omega
^{2}D_{i}^{2}D_{d}^{2}$) has the form
\begin{eqnarray}
A &=&-\lambda _{\omega }^{2}-\omega _{\nu d}^{2}\omega _{i1}\omega
_{i2}-\omega _{\nu i}^{2}\omega _{d1}\omega _{d2}+\omega _{i1}\omega
_{i2}\omega _{d1}\omega _{d2} \\
&&-\mu _{i}\mu _{d}\left( \omega _{i2}\omega _{d1}+\omega _{i1}\omega
_{d2}\right) .  \nonumber
\end{eqnarray}

The solutions for $v_{d1r,\theta }$ are obtained from Eqs. (D.1) and (D.2)
under substitution the index $i$ by the index $d$, and vice versa.

The solutions for the velocities $v_{i1z}$ and $v_{d1z}$ are the following:
\[
\lambda _{\omega }v_{i1z}=\omega _{\nu d}G_{i1z}-\mu _{d}G_{d1z},
\]
\begin{equation}
\lambda _{\omega }v_{d1z}=\omega _{\nu i}G_{d1z}-\mu _{i}G_{i1z}.
\end{equation}

\paragraph{APPENDIX E}

\smallskip
\paragraph{General solutions for the perturbed electron velocity in the
case $k_{z}\neq 0,$ $k_{r}=k_{\theta }=0$\\}
\renewcommand{\theequation}{E\arabic{equation}}
\setcounter{equation}{0} From the solutions given in the Appendix
B, we obtain the following expressions for the perturbed
velocities of the cold electrons:
\[
D_{e}v_{e1r}=-H_{e1r}-\mu _{ei}\omega _{\nu e}v_{i1r}-i\mu _{ei}\omega
_{e1}v_{i1\theta }-\mu _{ed}\omega _{\nu e}v_{d1r}-i\mu _{ed}\omega
_{e1}v_{d1\theta },
\]
\[
D_{e}v_{e1\theta }=-H_{e1\theta }+i\mu _{ei}\omega _{e2}v_{i1r}-\mu
_{ei}\omega _{\nu e}v_{i1\theta }+i\mu _{ed}\omega _{e2}v_{d1r}-\mu
_{ed}\omega _{\nu e}v_{d1\theta },
\]
\[
v_{e1z}=\frac{i}{\omega _{\nu e}}G_{e1z}+\frac{1}{\omega _{\nu e}\lambda
_{\omega }}\left[ \left( \mu _{i}\mu _{ed}-\omega _{\nu d}\mu _{ei}\right)
G_{i1z}+\left( \mu _{d}\mu _{ei}-\omega _{\nu i}\mu _{ed}\right)
G_{d1z}\right] ,
\]

\noindent where
\[
\mu _{ei}=\frac{\nu _{en}^{0}\nu _{ni}^{0}}{\omega _{\nu n}},\mu _{ed}=\frac{%
\nu _{en}^{0}\nu _{nd}^{0}}{\omega _{\nu n}}.
\]

\paragraph{APPENDIX F}

\paragraph{\noindent Values $W_{rr,\theta }$, $W_{\theta r,\theta }$, $%
W_{zrr,\theta ,z}$, $W_{z\theta r,\theta ,z}$, and $W_{zzr,\theta
,z}$ in the general form\\}
\renewcommand{\theequation}{F\arabic{equation}}
\setcounter{equation}{0} Values $W_{rr,\theta }$ and $W_{\theta
r,\theta }$ in the general form are the following:
\[
W_{rr}=W_{\theta \theta }=\lambda _{\omega }\left( q_{i}n_{i0}\omega _{\nu
d}-q_{d}n_{d0}\mu _{i}\right) +q_{e}n_{e0}\frac{\eta _{\omega }}{D_{e}}%
\alpha _{r1}+q_{e}n_{e0}i\omega _{\nu e}\frac{\omega _{ce}}{D_{e}\omega _{ci}%
}\triangle ,
\]
\[
W_{r\theta }=-W_{\theta r}=\omega _{\nu d}\omega _{ci}\left(
q_{i}n_{i0}\omega _{\nu d}-q_{d}n_{d0}\mu _{i}\right) +q_{e}n_{e0}\frac{\eta
_{\omega }}{D_{e}}\alpha _{r2}-q_{e}n_{e0}\frac{\omega _{ce}^{2}}{%
D_{e}\omega _{ci}}\triangle ,
\]

\noindent where
\begin{eqnarray*}
\alpha _{r1} &=&\lambda _{\omega }\omega _{\nu e}-i\omega _{\nu d}\omega
_{ce}\omega _{ci}, \\
\alpha _{r2} &=&i\lambda _{\omega }\omega _{ce}+\omega _{\nu d}\omega _{\nu
e}\omega _{ci}.
\end{eqnarray*}

\noindent The electron terms proportional to $\alpha _{r1,2}$ are connected
with the influence of the neutral-ion and neutral-dust collisions on the
collisions of electrons with neutrals.

Values $W_{zrr,\theta ,z}$, $W_{z\theta r,\theta ,z}$, and $W_{zzr,\theta
,z} $ in the general form are the following:
\begin{eqnarray*}
W_{zrr} &=&\beta _{1}\frac{n_{z}^{2}}{c^{2}}\left(
ib_{rz}v_{i0r}-v_{e0r}^{2}\right) ,W_{zr\theta }=\beta _{1}\frac{n_{z}^{2}}{%
c^{2}}\left( ib_{rz}v_{i0\theta }-v_{e0r}v_{e0\theta }\right) ,W_{zrz}=\beta
_{1}\frac{n_{z}}{c}\left( ib_{rz}-v_{e0r}\right) , \\
W_{z\theta r} &=&\beta _{1}\frac{n_{z}^{2}}{c^{2}}\left( ib_{\theta
z}v_{i0r}-v_{e0r}v_{e0\theta }\right) ,W_{z\theta \theta }=\beta _{1}\frac{%
n_{z}^{2}}{c^{2}}\left( ib_{\theta z}v_{i0\theta }-v_{e0\theta }^{2}\right)
,W_{z\theta z}=\beta _{1}\frac{n_{z}}{c}\left( ib_{\theta z}-v_{e0\theta
}\right) , \\
W_{zzr} &=&\beta _{1}\frac{n_{z}}{c}\left( ib_{zz}v_{i0r}-v_{e0r}\right)
,W_{zz\theta }=\beta _{1}\frac{n_{z}}{c}\left( ib_{zz}v_{i0\theta
}-v_{e0\theta }\right) ,W_{zzz}=\beta _{1}\left( ib_{zz}-1\right) .
\end{eqnarray*}
Here:
\begin{eqnarray*}
\beta _{1} &=&\frac{1}{4\pi i}\frac{\omega _{pe}^{2}}{\omega _{\nu e}}, \\
b_{r,\theta ,zz} &=&\frac{\omega _{\nu e}s_{r,\theta ,zz}}{\lambda _{\omega }%
}\frac{\omega _{ci}}{\omega _{ce}},
\end{eqnarray*}

\noindent where
\begin{eqnarray*}
s_{rz} &=&\frac{1}{q_{e}n_{e0}}\left( q_{i}n_{i0}\omega _{\nu
d}-q_{d}n_{d0}\mu _{i}\right) v_{i0r}+\frac{\eta _{\omega }}{D_{e}\omega
_{\nu e}}\left( \omega _{\nu e}\varepsilon _{er}+D_{e}v_{e0r}\right) , \\
s_{\theta z} &=&\frac{1}{q_{e}n_{e0}}\left( q_{i}n_{i0}\omega _{\nu
d}-q_{d}n_{d0}\mu _{i}\right) v_{i0\theta }+\frac{\eta _{\omega }}{%
D_{e}\omega _{\nu e}}\left( \omega _{\nu e}\varepsilon _{e\theta
}+D_{e}v_{e0\theta }\right) , \\
s_{zz} &=&\frac{1}{q_{e}n_{e0}}\left( q_{i}n_{i0}\omega _{\nu
d}-q_{d}n_{d0}\mu _{i}\right) +\frac{\eta _{\omega }}{\omega _{\nu e}}.
\end{eqnarray*}

\smallskip

\paragraph{\noindent REFERENCES\\}

\smallskip

\noindent \\
\noindent Adachi, I., Hayashi, C., \& Nakazawa, K. 1976, Prog. Theor. Phys.,
56, 1756

\noindent Balbus, S. A., \& Terquem, C. 2001, ApJ, 552, 235

\noindent Barge, P., \& Sommeria, J. 1995, A\&A, 295, L1

\noindent Barkan, A., D'Angelo, N., \& Merlino, R. L. 1994, Phys. Rev.
Lett., 73, 3093

\noindent Beckwith, S. V. W., \& Sargent, A. I. 1996, Nature, 383, 139

\noindent Bergin, E. A., Plume, R., Williams, J. P., \& Myers, P. C. 1999,
ApJ, 512, 724

\noindent Besla, G., \& Wu, Y. 2007, ApJ, 655, 528

\noindent Blaes, O. M., \& Balbus, S. A. 1994, ApJ, 421, 163

\noindent Carr, J. S., Tokunaga, A. T., \& Najita, J. 2004, ApJ, 603, 213

\noindent Chew, G. F., Goldberger, M. L., \& Low, F. E. 1956, Proc. R. Soc.
London, A 236, 112

\noindent Chow, V. W., Mendis, D. A., \& Rosenberg, M\textbf{. }1993, J.
Geophys. Res.,\ 98,\textbf{\ }19065

\noindent Cuzzi, J. N., Dobrovolskis, A. R., \& Champney, J. M. 1993,
Icarus, 106, 102

\noindent Desch, S. J. 2004, ApJ, 608, 509 \

\noindent Donati, J.-F., Paletou, F., Bouvier, J., \& Ferreira, J. 2005,
Nature, 438, 466

\noindent Draine, B. T., Roberge, W. G., and Dalgarno, A. 1983, ApJ, 264, 485

\noindent Ellis, T. A., \& Neff, J. S. 1991, Icarus, 91, 280

\noindent Fisher, D., \& Duerbeck, H. 1998, Hubble Revisited (Springer,
Berlin), p. 213

\noindent Fortov, V. E., Nefedov, A. P., Vaulina, O. S.\ et al. 1998, J.
Exp. Theor. Phys., 87, 1087

\noindent Goertz, C. K. 1989, Rev. Geophys.,\ 27, 271

\noindent Goldreich, P., \& Ward, W. R. 1973, ApJ, 183, 1051

\noindent Hartmann, L. 2000, Space Sci. Rev., 92, 55

\noindent Havnes, O., Brattli, A., Aslaksen, T. et al. 2001, Geophys. Res.
Lett., 28, 1419

\noindent Havnes, O., Tr\o im, J., Blix, T., Mortensen, W., Naesheim, L. I.,
Thrane, E., \& T\o nnesen, T. 1996, J. Geophys. Res.,\ 101, 10839

\noindent \noindent Hayashi, C. 1981, Suppl. Prog. Theor. Phys., 70, 3

\noindent Hayashi, C., Nakazawa, K., \& Nakagawa, Y. 1985, in Protostars and
Planets II, ed. D. C. Black \& M. S. Matthews (Tucson: Univ Arizona Press),
p. 1100

\noindent Hersant, F., Dubrulle, B., \& Hur\'{e}, J.-M. 2005, A\&A, 429, 531

\noindent Hor\'{a}nyi, M. 1996, Annu. Rev. Astron. Astrophys., 34,\textit{\ }%
383

\noindent Hor\'{a}nyi, M., \& Goertz, C. K. 1990, ApJ, 361, 155

\noindent Hutawarakorn, B., \& Cohen, R. J. 1999, MNRAS, 303, 845

\noindent ---------. 2005, MNRAS, 357, 338

\noindent Ilgner, M., \& Nelson, R. P. 2008, A\&A, 483, 815

\noindent \noindent Isella, A., Testi, L., \& Natta, A. 2006, A\&A, 451, 951

\noindent Jin, L. 1996, ApJ, 457, 798

\noindent Johansen, A., Andersen, A., \& Brandenburg, A. 2004, A\&A, 417, 361

\noindent Klahr, H. H., \& Bodenheimer, P. 2003, ApJ, 582, 869

\noindent Li, P. S., McKee, C. F., Klein, R. I., \& Fisher, R. T. 2008, ApJ,
684, 380

\noindent Lovelace, R., Li, H., Colgate, S., \& Nelson, A. 1999, ApJ, 513,
805

\noindent Matthews, L. S., Hayes, R. L., Freed, M. S., \& Hyde, T. W. 2007,
IEEE Trans. Plasma Sci., 35, 260

\noindent Melzer, A., Trottenberg, T., \& Piel, A. 1994, Phys. Lett. A,\
191,\ 301

\noindent Mendis, D. A., \& Rosenberg, M. 1994, Annu. Rev. Astron.
Astrophys.,\ 32\textbf{, }419

\noindent Merlino, R. L., \& Goree, J. A. 2004, Phys. Today, 57, 32

\noindent Meyer-Vernet, N. 1982, A\&A, 105, 98

\noindent Nekrasov, A. K. 2007, Phys. Plasmas, 14, 062107

\noindent ---------. 2008, Phys. Plasmas, 15\textbf{,} 032907

\noindent \noindent Norman, C., \& Heyvaerts, J. 1985, A\&A, 147, 247

\noindent Petersen, M. R., Julien, K., \& Stewart, G. R. 2007 a, ApJ, 658,
1236

\noindent Pinte, C., Fouchet, L., M\'{e}nard, F., Gonzalez, J.-F., \&
Duch\^{e}ne, G. 2007, A\&A, 469, 963

\noindent Pinto, C., Galli, D., \& Bacciotti, F. 2008, A\&A, 484, 1

\noindent Quanz, S. P., Apai, D., \& Henning, T. 2007, ApJ, 656, 287

\noindent Rotundi, A., Rietmeijer, F. J. M., Brucato, J. R. et al. 2000,
Planet. Space Sci., 48, 371

\noindent Safronov, V. S. 1969, Evolution of the Protoplanetary Cloud and
Formation of the Earth and the Planets (Moscow; Nauka)

\noindent Spitzer, Jr., L. 1978,\textbf{\ }Physical Processes in the
Interstellar Medium.\textit{\ }New York: Wiley

\noindent Thomas, H., Morfill, G. E., Demmel, V., Goree, J., Feuerbacher,
G., \& M\"{o}lmann, D. 1994, Phys. Rev. Lett., 73, 652

\noindent Tscharnuter, W. M., \& Gail, H.-P. 2007, A\&A, 463, 369

\noindent Wardle, M., \& Ng, C. 1999, MNRAS, 303, 239

\noindent Weidenschilling, S. J. 1995, Icarus, 116, 433

\noindent Whipple, E. C. 1981, Rep. Prog. Phys., 44, 1197

\noindent Yamoto, F., \& Sekiya, M. 2004, Icarus, 170, 180

\noindent Youdin, A. N., \& Goodman, J. 2005, ApJ, 620, 459

\noindent

\end{document}